\documentclass[twocolumn]{aastex631}

\usepackage{tabularx}
\usepackage{array}
\usepackage{bm}
\usepackage{sidecap}

\newcommand{\x}{\bm{x}}

\usepackage{xpatch}
\usepackage{hyperref}
\hypersetup{
	colorlinks=true,
	breaklinks=true,
	citecolor=blue,
	allcolors=blue,
	frenchlinks=true
}

\makeatletter
\xpatchcmd\NAT@citex
{%
	\@citea\NAT@hyper@{%
		\NAT@nmfmt{\NAT@nm}%
		\hyper@natlinkbreak{\NAT@aysep\NAT@spacechar}{\@citeb\@extra@b@citeb}%
		\NAT@date
	}%
}
{%
	\@citea
	\NAT@nmfmt{\NAT@nm}%
	\NAT@aysep\NAT@spacechar
	\NAT@hyper@{\NAT@date}%
}
{}{}
\xpatchcmd\NAT@citex
{%
	\@citea\NAT@hyper@{%
		\NAT@nmfmt{\NAT@nm}%
		\hyper@natlinkbreak{\NAT@spacechar\NAT@@open\if*#1*\else#1\NAT@spacechar\fi}%
		{\@citeb\@extra@b@citeb}%
		\NAT@date
	}%
}
{
	\@citea
	\NAT@nmfmt{\NAT@nm}%
	\NAT@spacechar\NAT@@open\if*#1*\else#1\NAT@spacechar\fi
	\NAT@hyper@{\NAT@date}%
}
{}{}
\makeatother

\accepted{June 12, 2025}

\begin{document}

\title{Discovery of Diffuse Radio Emission in a Massive $z = 1.709$ Cool Core Cluster: \\A Candidate Radio Mini-Halo}

\shorttitle{A Radio Mini-Halo at $z=1.709$}
\shortauthors{Hlavacek-Larrondo et al.}

\correspondingauthor{Julie Hlavacek-Larrondo}
\email{j.larrondo@umontreal.ca}

\author[0000-0002-0786-7307]{Julie HLavacek-Larrondo}
\affiliation{D\'{e}partement de Physique, Universit\'{e} de Montr\'{e}al, Succ. Centre-Ville, 
Montr\'{e}al, Qu\'{e}bec, H3C 3J7, Canada}

\author{Roland Timmerman}
\affiliation{Centre for Extragalactic Astronomy, Department of Physics, Durham University, South Road, Durham, DH1 3LE, United Kingdom}
\affiliation{Institute for Computational Cosmology, Department of Physics, Durham University, South Road, Durham DH1 3LE, United Kingdom}

\author[0000-0002-7275-3998]{Christoph Pfrommer}
\affiliation{Leibniz-Institut f\"{u}r Astrophysik Potsdam (AIP), 
    An der Sternwarte 16, D-14482 Potsdam, Germany}

\author{Erik Osinga}
\affiliation{Dunlap Institute for Astronomy and Astrophysics, University of Toronto, 50 St. George Street, Toronto, ON M5S 3H4, Canada}

\author{Larissa Tevlin}
\affiliation{Leibniz-Institut f\"{u}r Astrophysik Potsdam (AIP), 
    An der Sternwarte 16, D-14482 Potsdam, Germany}
\affiliation{Institut für Physik und Astronomie, Universität Potsdam, Karl-Liebknecht-Str. 24/25, 14476 Potsdam, Germany}

\author{Tracy M. A. Webb}
\affiliation{Department of Physics, McGill Space Institute, McGill University, 3600 rue University, Montr´eal, Qu´ebec, Canada, H3A 2T8}

\author{Natalia Martorella}
\affiliation{Department of Physics, McGill Space Institute, McGill University, 3600 rue University, Montr´eal, Qu´ebec, Canada, H3A 2T8}

\author{Xiaoyuan Zhang}
\affiliation{Max-Planck-Institut fur Extraterrestrische Physik, Giessenbachstrasse, 85748, Garching, Germany}

\author{Reinout van Weeren}
\affiliation{Leiden Observatory, Leiden University, PO Box 9513, 2300 RA Leiden, The Netherlands}

\author{Hyunseop Choi}
\affiliation{D\'{e}partement de Physique, Universit\'{e} de Montr\'{e}al, Succ. Centre-Ville, 
Montr\'{e}al, Qu\'{e}bec, H3C 3J7, Canada}

\author{Gabriella Di Gennaro}
\affiliation{Instituto di Radioastronomia, via P. Gobetti 101, 40129 Bologna, Italy}

\author{Marie-Lou Gendron-Marsolais}
\affiliation{Département de physique, de génie physique et d’optique, Université Laval, Québec (QC), G1V 0A6, Canada}

\author[0000-0003-2001-1076]{Carter Rhea}
\affiliation{D\'{e}partement de Physique, Universit\'{e} de Montr\'{e}al, Succ. Centre-Ville, 
Montr\'{e}al, Qu\'{e}bec, H3C 3J7, Canada}
\affiliation{Centre de Recherche en Astronomie du Qu\'{e}bec (CRAQ)}
\affiliation{Dragonfly Focused Research Organization, 150 Washington Avenue, Santa Fe, 87501, NM, USA}

\begin{abstract}
Clusters of galaxies host spectacular diffuse radio sources, extending over scales from 100 kpc to several Mpcs. These sources, with extremely faint surface brightness ($\mu$Jy/arcsec$^2$ level), are not tied to individual galaxies but trace synchrotron emission from large-scale magnetic fields and relativistic particles within the intracluster environment. Here, we report the discovery of a candidate radio mini-halo in SpARCS104922.6+564032.5, the most distant cool-core galaxy cluster identified to date at $z=1.709$, using deep LOFAR 120–168 MHz observations. We show that this emission originates from diffuse cluster-associated processes rather than unresolved AGN or star-forming galaxies. The diffuse radio emission coincides spatially with the X-ray emission of the hot intracluster medium and has a radio power of $P_{\rm 150~MHz}=49.8^{+14.7}_{-11.7} \times10^{24}$ W Hz$^{-1}$, exhibiting striking similarities to low-redshift radio mini-halos. This discovery doubles the redshift of previously known mini-halos, challenging models of inverse Compton losses and indicating the presence of strong magnetic fields, enhanced turbulence in high-redshift clusters, or active hadronic processes that require a cosmic ray-to-thermal energy ratio of 0.07 within 200~kpc, assuming a clumped distribution with spatial correlations among the gas, cosmic rays, and magnetic field that partially compensate for cosmological redshift dimming. It further implies that magnetic fields are efficiently amplified to $\sim$$10~\mu$G levels within a Mpc$^3$ volume during the epoch of cluster formation before $z\sim2$. These findings provide critical insights into high-redshift cluster physics and emphasize the transformative potential of next-generation radio surveys, such as those with the SKA and ngVLA, in exploring the early evolution of galaxy clusters.

\end{abstract}


\section{Introduction} \label{sec:intro}

Galaxy clusters are the largest gravitationally bound structures, consisting of hundreds to thousands of galaxies, hot X-ray emitting gas (the Intracluster Medium, ICM), and dark matter. Forming at cosmic filament intersections, they serve as laboratories for studying large-scale structure formation and evolution. They also host spectacular diffuse radio sources extending over $\sim$100 kpc to Mpcs, with extremely faint surface brightness ($\mu$Jy/\arcsec$^2$ at GHz frequencies). These sources are not tied to individual galaxies but trace synchrotron emission from large-scale magnetic fields and energetic particles, offering insight into particle transport, reacceleration physics, and magnetic fields \citep[see reviews by][]{Brunetti_2014,Ruszkowski_2023}.

Advances with the Jansky Very Large Array \citep[JVLA; e.g.][]{Perley2011}, MeerKAT \citep[e.g.][]{Jonas2016}, the upgraded Giant Metrewave Radio Telescope \citep[uGMRT; e.g.][]{Gupta2017}, and LOFAR \citep[e.g.][]{vanHaarlem2013}, along with low-frequency surveys like GLEAM \citep{wayth2015gleam,hurley2017galactic} and the LOFAR Two Metre Sky Survey \citep[LoTSS;][]{shimwell2017lofar,Shimwell_2019}, have expanded our knowledge of these sources. They are generally categorized into four groups: (1) radio mini-halos in cool core clusters, (2) giant radio halos in disturbed clusters, (3) radio relics, and (4) revived AGN fossil plasmas \citep[see review by][]{van_Weeren_2019}.

Radio mini-halos have been identified in $\sim$40 cool core clusters \citep[e.g.][]{Giacintucci_2013,begin2023extended,knowles2022meerkat,PE2020}, yet their origin remains unclear. The radiative timescale of AGN jets in the central galaxy is too short for electrons to reach the observed extents. Possible explanations include the reacceleration of seed cosmic rays (CRs) by turbulence from AGN outbursts \citep{Bravi2016,RL2020}, shocks \citep[][]{Bonafede2023}, or minor mergers driving sloshing-induced turbulence \citep{gitti2004particle,Mazzotta2008,Biava2024}. Alternatively, the hadronic model suggests a continuous injection of electrons via inelastic collisions of high-energy protons with ICM protons \citep[e.g.][]{Pfrommer2004}.

Recent high-resolution, multi-frequency studies of nearby radio mini-halos have uncovered correlations with X-ray features in the ICM \citep[e.g.][]{gendron2017deep,GM2021,giacintucci2019expanding,Biava2024,vanweeren2024}, revealing fundamental links between thermal and non-thermal cluster particles. LOFAR studies also show that some clusters hosting radio mini-halos exhibit more diffuse, large-scale emission resembling giant radio halos \citep[e.g.][]{Savini2018,Savini2019,Cassano2023,vanweeren2024}.

Enhanced sensitivity from new radio facilities has led to the discovery of several high-redshift halos, including PSZ2 G099.86+58.45 \citep[$z=0.616$;][]{cassano2019lofar}, PLCKG147.32–16.59 \citep[$z=0.65$;][]{van_Weeren_2014}, El Gordo \citep[$z=0.87$;][]{Lindner_2014}, PSZ2 G091.83+26.11 \citep[$z=0.822$;][]{DiGennaro2021,DiGennaro2023}, PSZ2 G160.83+81.66 \citep[$z=0.888$;][]{Di_Gennaro_2021}, and the most distant to date, ACT-CL J0329.2-2330 \citep[$z=1.23$;][]{Sikhosona2024}. The most distant known radio mini-halos are in the Phoenix cluster \citep[$z=0.596$;][]{van_Weeren_2014_phoenix,Raja_2020,Timmerman_2021} and ACT-CL J0022.2-0036 \citep[$z=0.8$;][]{Knowles_2019}. These findings suggest ICM magnetic fields are rapidly amplified in early cluster formation \citep{Tevlin_2024} to counteract inverse Compton (IC) losses \citep[see also][]{DiGennaro2025}, which scale as $\propto (1+z)^4$ and make diffuse radio emission at high redshifts fainter \citep{En_lin_2002,Cassano_2006,cassano2019lofar,DiGennaro2020,DiGennaro2021}. The detection of steep-spectrum radio halos ($\alpha<-1.5$, where $S\propto\nu^\alpha$) highlights the critical role of low-frequency telescopes like LOFAR in exploring these structures, and further detections at $z>1$ could offer key insights into magnetic field evolution.

\begin{figure*}
\centering
\includegraphics[width = 1\textwidth]{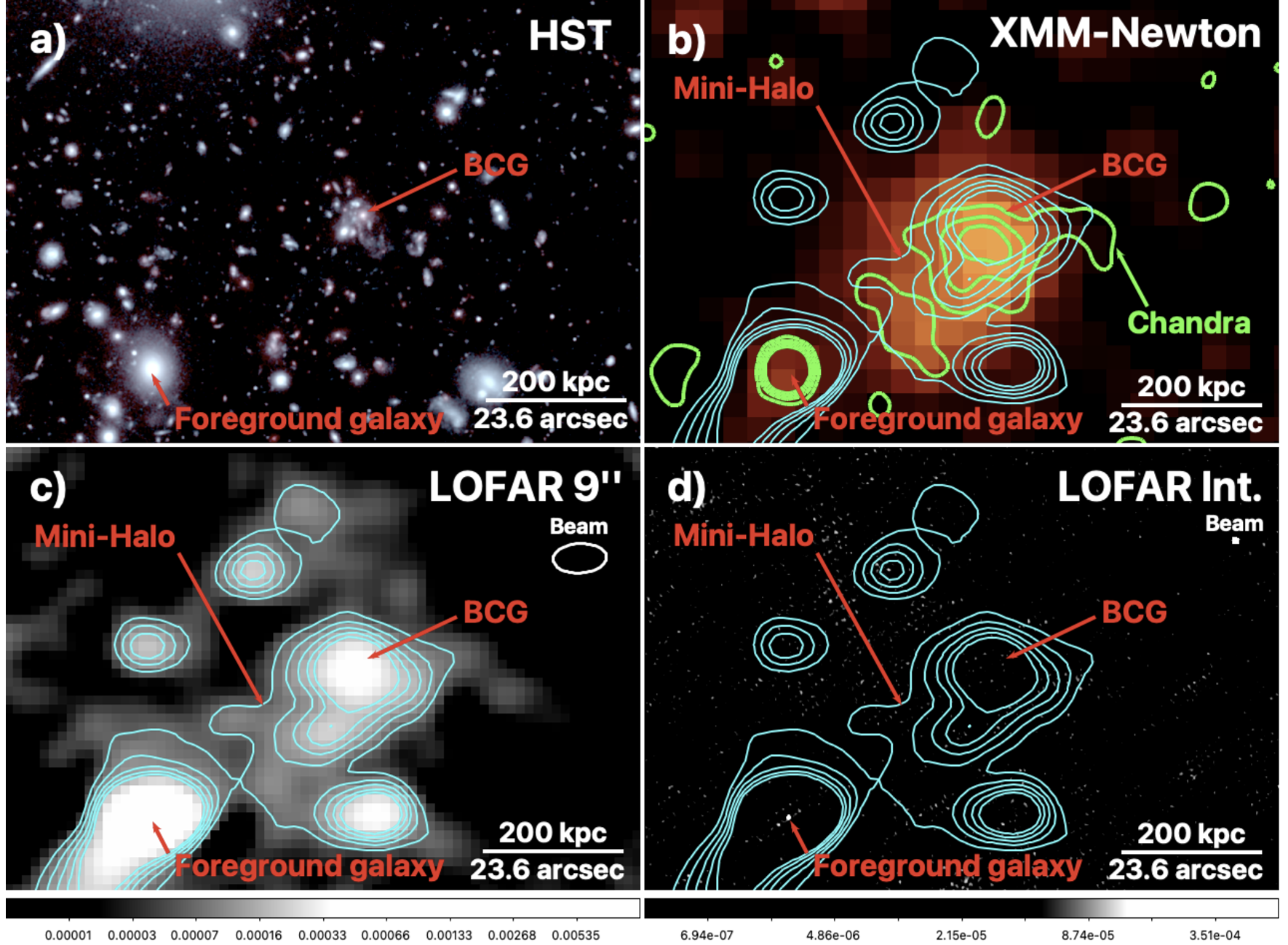}
\caption{Large-scale image of the $z=1.709$ cluster of galaxies SpARCS1049+56. Top-Left (a): Composite HST image (F105W in blue and green; F160W in red) of the cluster. We highlight the location of the BCG and a foreground $z\sim0.35$ galaxy located to the south-east, labeled S3 in Fig. \ref{fig:hadronic_model} and discussed in Appendix \ref{appB}. Top-right (b): XMM-Newton 0.5-2.0 keV X-ray image of the cluster. The image has been smoothed with a Gaussian filter of $\sigma$=2 pixels. Overplotted are the Chandra X-ray contours at 2, 3, 4 and 5 $\sigma_{\rm rms}$ in green, which show the compact core of the cluster. The cyan contours show the diffuse extended LOFAR emission and are the same as the bottom-left plot. Bottom-left (c): 144 MHz LOFAR 9\arcsec\ image (Dutch stations only) with a beam size of 8.9\arcsec$\times$4.9\arcsec. Contours are 2, 3, 4, 5 and 6$\sigma_{\rm rms}$ where $\sigma_{\rm rms}=35\mu$Jy/beam. This image reveals a diffuse extended halo. Bottom-right (d): 144 MHz LOFAR image, obtained with the International LOFAR Telescope, with a beam size of 0.39\arcsec$\times$0.22\arcsec. The contours are the same as the bottom-left plot. Here, the only source detected is the AGN in the foreground $z\sim0.35$ galaxy, implying that the emission detected in the bottom-left is of diffuse nature. Color bars are included for radio images in units are Jy/beam. }
\label{fig:largescale}
\end{figure*}

Here, we present the discovery of a candidate radio mini-halo in the highest redshift cool core galaxy cluster known, SpARCS104922.6+564032.5 (SpARCS1049+56), at $z=1.709$ \citep[Fig.~\ref{fig:largescale};][]{webb_extreme_2015,webb_detection_2017,Hlavacek_Larrondo_2020}. Initially reported by \citet{Osinga_2021}, the emission was thought to arise from unresolved AGN. We demonstrate instead that it likely originates from diffuse cluster-scale processes, not compact sources such as AGN or star-forming galaxies. This is puzzling given expected IC losses and suggests new physics governing high-redshift clusters. Section \ref{sec:obs} presents the observations, Section \ref{res} the results, and Section \ref{sec:disc} their implications. We adopt $H_0=70$ km s$^{-1}$ Mpc$^{-1}$, $\Omega_{\rm m}=0.3$, and $\Omega_{\rm \Lambda}=0.7$, with 8.463 kpc per \arcsec at $z=1.709$. Errors are 1$\sigma$ unless stated otherwise.

\section{Radio Observations} \label{sec:obs}

\begin{table*}[ht]
\caption{Properties of the LOFAR images taken at 120–168 MHz (averaging 144 MHz) at both low spatial resolution (using Dutch stations only) and high spatial resolution (including all international baselines). 1) Name given to the resulting image and shown in other figures. 2) Stations used. 3) Total integration time in hours (hrs). 4) If applicable, the $uv$ taper used to produce the image in units of kpc at the cluster redshift. 5) Beam size. 6) Beam position angle (BPA). 7) Resulting noise level near the cluster. } 
\centering
\begin{tabular*}{\textwidth}{@{\extracolsep{\fill}}lcccccc}
       \hline
       Name  & Stations & Integration time & $uv$ taper & Bmaj$\times$Bmin & BPA & RMS \\
       (1) & (2) & (3) & (4) & (5) & (6) & (7) \\
       \hline
        LOFAR 9\arcsec  & Dutch & 8 hrs & ... & 8.9\arcsec $\times$ 4.9\arcsec & 93\(^\circ\) & 35 \(\mathrm{\mu}\)Jy beam\(^{-1}\) \\
       LOFAR 6\arcsec & Dutch & 8 hrs & ... & 6.0\arcsec $\times$ 6.0\arcsec & -5.9\(^\circ\) & 33 \(\mathrm{\mu}\)Jy beam\(^{-1}\) \\
       LOFAR Taper50 & Dutch & 8 hrs & 50 & 13.8\arcsec $\times$ 8.6\arcsec & 93\(^\circ\) & 53 \(\mathrm{\mu}\)Jy beam\(^{-1}\) \\
       LOFAR Sub & Dutch & 8 hrs & 50 & 13.8\arcsec $\times$ 8.6\arcsec & 93\(^\circ\) & 43 \(\mathrm{\mu}\)Jy beam\(^{-1}\) \\
       LOFAR Int. & All & 8 hrs & ... & 0.39\arcsec $\times$ 0.22\arcsec & 17\(^\circ\) & 61 \(\mathrm{\mu}\)Jy beam\(^{-1}\) \\
\hline

    \end{tabular*}
    \label{tab}
\end{table*}

\subsection{LOFAR observations (Dutch stations)}

SpARCS1049+56 was observed using LOFAR (Dutch stations only) as part of the LoTSS Deep Fields survey \citep{Tasse2021,Sabater2021,Kondapally2021,Duncan2021}. This survey covers three regions with extensive multi-wavelength data: the European Large-Area ISO Survey-North 1 \citep[ELAIS-N1;][]{Oliver2000}, Boötes \citep{Jannuzi1999}, and the Lockman Hole \citep{Lockman1986}, spanning over 50 deg$^2$. It aims for noise levels as low as $10-15\mu$Jy beam$^{-1}$ \citep{Tasse2021}, with SpARCS1049+56 located in the Lockman Hole field. The first data release, based on 80 hours of observation, achieved a central noise level of $\sim$22 $\mu$Jy beam$^{-1}$, with details on observations and data reduction in \citet{Tasse2021}.

LOFAR observations of SpARCS1049+56, first presented by \citet{Osinga_2021}, revealed extended cluster-associated emission (see their Fig. 8). The bottom-left panel of Fig.~\ref{fig:largescale} displays the 120–168 MHz image (average of 144 MHz) from \cite{Osinga_2021}, captured with the LOFAR High Band Antenna (HBA). This image was generated by reprocessing the LoTSS Deep Fields data, specifically optimized for the cluster’s location, resulting in a $8.9\arcsec\times5.9\arcsec$ beam size (hereafter referred to as the LOFAR 9\arcsec\ image). This resolution is ideal for capturing the extent of the diffuse component, with a noise level near the cluster of $\sigma_{\rm rms}=35\mu$Jy beam$^{-1}$. We also produced a higher-resolution image with a beam of $6.0\arcsec\times6.0\arcsec$ (hereafter the LOFAR 6\arcsec\ image), shown in Figs.~\ref{fig:zoom}, \ref{fig:hadronic_model}, \ref{fig:supp1} and \ref{fig:supp2}. This improved resolution better isolates diffuse emission from compact sources, making it the primary dataset for our cluster-scale analysis (see Section \ref{res}).

To further enhance visibility of the diffuse emission, we also reprocessed the data using different $uv$-tapers and generated a point-source-subtracted image \citep[see][]{Osinga_2021}. While these images reinforce the presence of extended emission, larger beam sizes complicate separation from nearby sources. In Fig.~\ref{fig:supp1} of Appendix \ref{appA}, we present a 50 kpc $uv$-tapered image and the point-source-subtracted image. Specifications of all images are detailed in Table \ref{tab}.

\subsection{LOFAR observations (all stations)}

To assess whether the extended emission in SpARCS1049+56 is diffuse or from unresolved AGN, we reprocessed archival LOFAR data (Project: \textsc{LT16\_005}, SAS id: 876618) and the primary calibrator 3C,196 (SAS id: 876616) observed on November 17, 2022. Both were recorded with full linear polarization, 1-second time integrations, and a 120–168~MHz bandwidth with 12.2~kHz channels.

Calibration was performed using LINC \citep{Williams2016,vanWeeren2016,deGasperin2019}, with radio-frequency interference removed via \textsc{AOFlagger} \citep{Offringa13, Offringa15}. The process included bandpass, clock offset, and polarization alignment corrections based on a model of the primary calibrator, applied solely to the Dutch stations of the target observation. Phase calibration was then refined using a sky model from the TIFR Giant Metrewave Radio Telescope Sky Survey \citep[TGSS;][]{Intema2017}, followed by a single round of self-calibration.

The calibration was then extended to the international LOFAR stations using the LOFAR-VLBI pipeline \citep{morabito21}. First, solutions from the primary calibrator were applied to international baselines, followed by phase corrections accounting for clock offsets and ionospheric delays using bright calibrators in the field. Due to strong ionospheric perturbations, we derived phase corrections from five bright sources near SpARCS1049+56 instead of a distant in-field calibrator. These sources were phase-shifted, divided by model visibilities to approximate point sources, and stacked to create a combined dataset for calibration.

\begin{figure*}
\centering
\includegraphics[width = 1\textwidth]{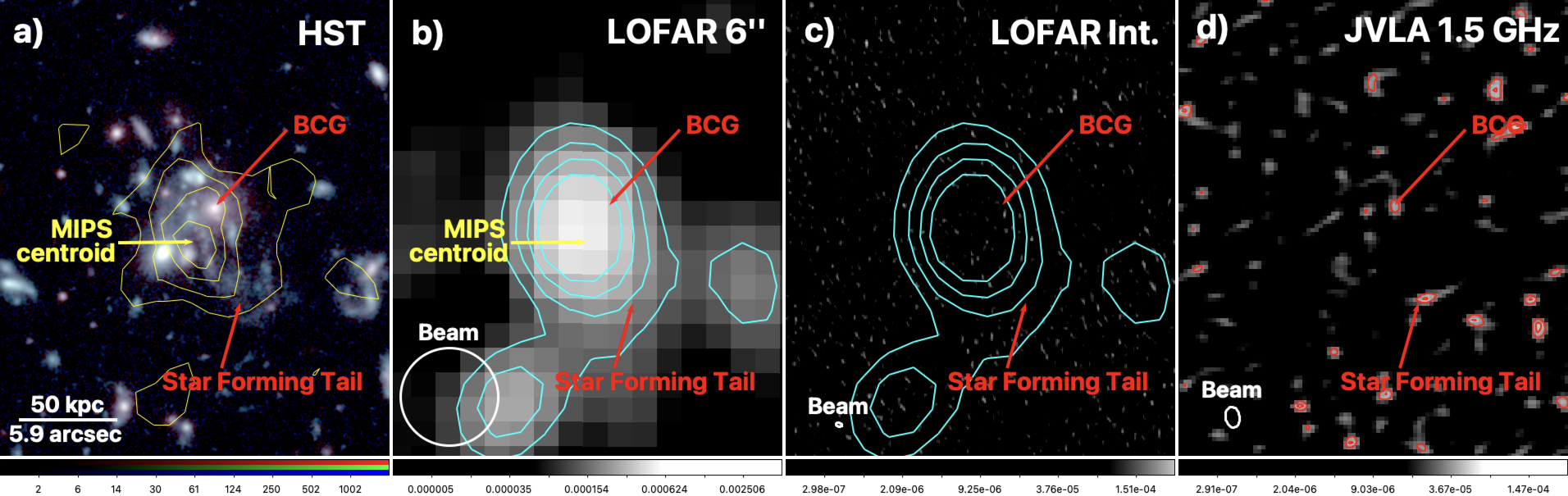}
\caption{Zoom-in on the star-forming region of SpARCS1049+56, labeled as S2 in Fig. \ref{fig:hadronic_model}. 
Left (a): Composite HST image (F105W in blue and green; F160W in red), highlighting the BCG and the star-forming tidal-like tail. Overlaid yellow contours represent the Spitzer 24 $\mu$m emission from \cite{webb_extreme_2015} at 2, 3, 4 and 5$\sigma_{\rm rms}$. The centroid of the MIPS emission is highlighted, and the emission appears slightly extended along the star-forming tail. 
Middle-left (b): 144 MHz LOFAR 6\arcsec\/ image with a beam of 6.0\arcsec$\times$6.0\arcsec. Contours are 4, 6, 8 and 10$\sigma_{\rm rms}$, where $\sigma_{\rm rms}=33\mu$Jy/beam. The centroid of the MIPS emission coincides with the peak of the radio emission (labeled S2 in Fig. \ref{fig:hadronic_model}). 
Middle-right (c): 144 MHz LOFAR image obtained with the International LOFAR Telescope, with a beam of 0.39\arcsec$\times$0.22\arcsec, using the same low-resolution contours as in the middle-left panel. No source is detected, indicating that the LOFAR emission in the middle-left panel is of diffuse origin. 
Right (d): JVLA 1.5 GHz image from \cite{trudeau_multiwavelength_2019}, with contours at 3, 4 and 5$\sigma_{\rm rms}$, where $\sigma_{\rm rms}=10.5\mu$Jy/beam. Color bars are included; radio image units are Jy/beam.}
\label{fig:zoom}
\end{figure*}

To minimize interference, LOFAR's core stations were digitally phased into a single superstation, data were averaged to 16-second time integrations and 195~kHz spectral channels, and only baselines longer than 80~k\(\mathrm{\lambda}\) were used. After applying the final phase corrections, the data were phase-shifted to SpARCS1049+56 to produce the final calibrated dataset (see Figs.~\ref{fig:largescale} and \ref{fig:zoom}). The resulting image has a beam size of $0.39\arcsec\times0.22\arcsec$ and a noise level of $\sigma_{\rm rms}=61\mu$Jy beam$^{-1}$. Details are in Table \ref{tab}.

\section{Results}\label{res}

\subsection{Radio emission from the cluster}\label{res:cl}

\citet{Osinga_2021} first detected diffuse radio emission in SpARCS1049+56, interpreting it as AGN-related but noting difficulties in separating compact sources from the diffuse component, particularly given the cluster’s high redshift. In this work, we present archival LOFAR International observations at the same frequency as \cite{Osinga_2021} but at a much higher spatial resolution. If the emission were due to unresolved AGNs, it should be evident in these data. However, as shown in the bottom-right panel of Figure~\ref{fig:largescale}, no compact source is associated with the cluster, not even the Brightest Cluster Galaxy (BCG). The only detected compact source lies within a foreground galaxy at $z\sim0.35$ to the southeast, labeled as S3 in Fig. \ref{fig:hadronic_model} and further discussed in Appendix \ref{appB}.

Consequently, the absence of point-like emission in the high-resolution LOFAR International image rules out active AGNs as the source, confirming that the emission in \cite{Osinga_2021} must be diffuse. This aligns with \citet{trudeau_multiwavelength_2019}, who conducted deep JVLA imaging in L, C, and X bands (1.4 GHz, 6 GHz, and 10 GHz). Their study detected mainly a faint AGN at the BCG’s location, along with the foreground $z\sim0.35$ galaxy. 

To further characterize the nature of the diffuse emission and assess whether it could be resolved out by the LOFAR International image, we estimated the minimum angular size that the brightest component must reach for its surface brightness to fall below the detection threshold. Specifically, using a threshold of $3\times\sigma_{\rm rms}=0.18$ mJy/beam, we find that the bright \(0.54\) mJy emission detected in the LOFAR 6'' image must be extended at least 3 LOFAR International beams to be resolved out in the LOFAR International image.

We note that the LOFAR data alone cannot exclude a scenario where past AGN activity allowed electrons to diffuse, forming the large-scale radio emission. If so, intermediate-scale emission between 0.3\arcsec and 0.6\arcsec\ may be missing. We emphasize this in Section 4.1 when discussing the limitations of the data, but note that both the deep JVLA L-band image (beam size: 1.3\arcsec$\times$0.9\arcsec) from \citet[][]{trudeau_multiwavelength_2019} and the LOFAR data with a \textit{uv}-taper applied (beam size: 3.9\arcsec$\times$0.8\arcsec) detect no such emission, further supporting the diffuse nature of the emission detected in \citet[][]{Osinga_2021}\footnote{The JVLA L-band data in \citet[][]{trudeau_multiwavelength_2019} were taken in A-configuration, probing intermediate scales of $\sim$1\arcsec, but not diffuse emission on larger scales seen with LOFAR (i.e. $\sim$6\arcsec\ or hundreds of kpc). Thus, the non-detection of the candidate radio mini-halo in the current JVLA data is expected. JVLA L-band data in B-configuration would be needed to detect emission on similar scales as LOFAR.}.

In summary, Fig.~\ref{fig:largescale} reveals two diffuse components in the LOFAR emission: a bright region near the BCG extending to $\sim$80 kpc (labeled S2 in Fig. \ref{fig:hadronic_model} and highlighted in Fig.~\ref{fig:zoom}) and a fainter component to the southeast, spanning $\sim$350 kpc or 41\arcsec\ at the cluster’s redshift (labeled as S1 in Fig. \ref{fig:hadronic_model}).

In Section \ref{res:sf}, we argue that the first component (S2) originates from diffuse radio emission driven by the intense starburst in SpARCS1049+56's core, while the second (S1) cannot be attributed to star formation in the BCG or any cluster member. This suggests an alternative origin for the $\sim$350 kpc diffuse emission (S1). The bottom-left panel of Fig.~\ref{fig:largescale} shows a radio-bright foreground galaxy at $z\sim0.35$ detected at 144 MHz with LOFAR (labeled as S3 in Fig. \ref{fig:hadronic_model}). To its southeast, extended jet-like radio emission is visible. If this were a radio jet, the diffuse emission (S1) in SpARCS1049+56 might be its northwestern counterpart. However, Appendix \ref{appB} shows this is not the case: the apparent jet to the southeast is actually several distinct AGNs, including the AGN in the $z\sim0.35$ galaxy. Thus, no radio jet extends southeastward, ruling out a jet counterpart as the source of the diffuse emission.

In conclusion, the $\sim$350 kpc LOFAR emission (S1) is not due to unresolved AGN activity, star formation, or a radio jet. Its faint, large-scale nature, central location within the cluster, and coincidence with X-ray emission (Fig.~\ref{fig:largescale}) suggest it could be a radio mini-halo. We explore the implications in Section \ref{sec:disc}.

\subsection{Radio emission from star formation}\label{res:sf}

Figure~\ref{fig:zoom} presents a zoom-in of the cluster core, highlighting the radio source labeled S2 in Fig. \ref{fig:hadronic_model}. S2 is slightly elongated north-south, spanning $\sim$$9\arcsec\times13\arcsec$ ($\sim$75 kpc $\times100$ kpc). It is offset south of the BCG, aligning instead with the $860\pm130$ M$_{\rm \odot}$ yr$^{-1}$ starburst in SpARCS1049+56’s core, following the tidal-like tail seen in the HST image \citep[Fig.~\ref{fig:zoom};][]{webb_extreme_2015,webb_star_2015,webb_detection_2017}. Notably, the peak of the radio emission of S2 aligns with the centroid of the Spitzer 24-micron emission, tracing the starburst. The middle-right panel of Fig.~\ref{fig:zoom} shows that this emission is undetected in high-resolution LOFAR data, confirming its diffuse nature rather than from compact sources like AGNs. Thus, the $\sim$75 kpc $\times100$ kpc extended radio source in the core is likely the radio counterpart of the starburst. A detailed analysis of its radio power relative to star formation rate will be presented in Webb et al. (in prep).

As discussed earlier, Fig.~\ref{fig:largescale} also reveals a faint, extended radio component spanning $\sim$350 kpc (labeled S1 in Fig. \ref{fig:hadronic_model}), interpreted as a candidate radio mini-halo, in addition to the $\sim$75 kpc $\times100$ kpc starburst-associated component. S1 cannot be from diffuse star formation in cluster members or confusion in a crowded field. Apart from the quiescent BCG and the intense starburst, no spectroscopic or photometric-redshift cluster members coincide with the mini-halo. While some small, faint galaxies in the deep HST image align spatially with the mini-halo, their absence in the Spitzer 24$\mu$m image suggests they lack sufficient star formation to contribute to the mini-halo.

\section{Discussion and Implications} \label{sec:disc}

\subsection{Limitations with the current observations}

We emphasize that the candidate radio mini-halo is very faint. As discussed in Section~\ref{res:cl}, some emission on intermediate scales ($\sim0.3$–$6$ arcsec) may still be missed, despite no clear evidence in the deep JVLA L-band (beam: 1.3\arcsec$\times$0.9\arcsec) and LOFAR \textit{uv}-tapered data (beam: 3.9\arcsec$\times$0.8\arcsec). Additional observations, such as JVLA P-band (350 MHz) or LOFAR Low Band Antenna, are necessary to identify potential missing emission and confirm the radio mini-halo origin. A detection at another frequency would also allow for spectral index measurements, providing a key test of consistency with the mini-halo population \citep[$\alpha \sim -1.15$; e.g.,][]{giacintucci2019expanding}. Indeed, it is not possible to extract a reliable spectral index using the LOFAR HBA detection alone \citep[see e.g.][]{shimwell2022}.

\subsection{Properties of the radio mini-halo}

Assuming that the candidate radio mini-halo is indeed a radio mini-halo, we note that its properties are remarkably similar to those of other known radio mini-halos \citep[e.g.,][]{Giacintucci_2013,begin2023extended,knowles2022meerkat}. Specifically, the source spans a size of $\sim$350 kpc, consistent with the typical sizes of radio mini-halos found at low redshifts \citep[][]{RL2020}. Furthermore, the emission is centered toward the south-east, mirroring the orientation of the X-ray emission from the cluster (see Fig.~\ref{fig:largescale}), although we caution that the limited resolution and sensitivity of LOFAR prevent a detailed morphological comparison. SpARCS1049+56 is the highest-redshift known cool-core cluster \citep[see][]{Hlavacek_Larrondo_2020}, with a mass of $M_{\rm 200,c}=(3.5\pm1.2)\times10^{14}$ M$_\odot$ \citep[][]{Finner_2020}, placing it among the most massive clusters at this redshift. As demonstrated by \citet[][]{giacintucci2019expanding}, radio mini-halos are also predominantly found in massive cool-core clusters.

Here, we present the corresponding radio power at 1.4 GHz, derived as follows. The flux density of the radio mini-halo was first estimated using the \textsc{Halo-FDCA} package \citep{boxelaar21}. We fitted a circular Gaussian profile to the radio mini-halo using the high-resolution LOFAR 6\arcsec{} image. By applying a mask to the image, this fit only considered emission from the radio mini-halo, separating its emission from that of the starbursting core located near the BCG (labeled S2 in Fig. \ref{fig:hadronic_model}) and all nearby AGN (i.e. all sources highlighted in magenta in Fig. \ref{fig:hadronic_model}). A 10$\%$ uncertainty on the flux density scale was included in the estimate, yielding a 144 MHz flux density of $S_{\rm 144 MHz}=2.29^{+0.56}_{-0.38}$ mJy. This value was then converted to the expected 1.4 GHz flux density using a typical mini-halo spectral index, $\alpha=-1.15\pm0.15$ \citep{giacintucci2019expanding}. The corresponding radio power was calculated using $P_{\rm 1.4 GHz}=4\pi S_{\rm 1.4 GHz}D_\mathrm{lum}^2(1+z)^{-(\alpha+1)}$, where $D_\mathrm{lum}$ is the luminosity distance, with uncertainties propagated throughout, yielding $P_{\rm 1.4 GHz}=3.82^{+0.94}_{-0.64} \times10^{24}$ W Hz$^{-1}$. At 150 MHz, the corresponding radio power is $P_{\rm 150 MHz}=49.8^{+14.7}_{-11.7} \times10^{24}$ W Hz$^{-1}$. These values are well within the range of other known radio mini-halos.

To provide context for this result, Fig.~\ref{fig:plot} compares the radio power of the mini-halo in SpARCS1049+56 to other radio mini-halos as a function of X-ray luminosity, measured within the central 600 kpc \citep[sample from][]{RL2020}. The X-ray luminosity of SpARCS1049+56 was determined using 100 ks \emph{XMM-Newton} observations of the cluster (ObsID 0940830601). The original data files of the European Photon Imaging Camera (EPIC) were processed using the pipeline described in \citet{Zhang2023}. The flare-removed clean event files have exposures of 60 ks. 
We extract the central 600 kpc region spectra and model them with a single temperature APEC model to calculate the rest-frame $0.1$--$2.4$ keV luminosity, assuming a fixed metallicity of 0.3\,Z$_\odot$. This analysis yielded an X-ray luminosity within $r = 600$ kpc of $L_{\rm X} = (2.4\pm0.3) \times 10^{44}$ erg s$^{-1}$. Figure~\ref{fig:plot} shows that the diffuse radio emission in SpARCS1049+56 is consistent with the general population of radio mini-halos observed at low redshift ($z\sim0.2$), although it lies toward the radio brighter end of the distribution.

\subsection{Implications for cluster astrophysics}
\label{sec:astrophysics}

\begin{figure}
\centering
\includegraphics[width = \columnwidth]{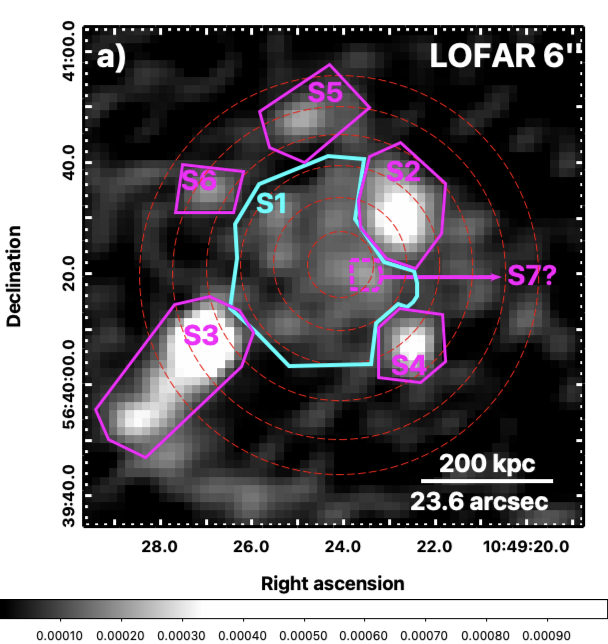}
\includegraphics[width = \columnwidth]{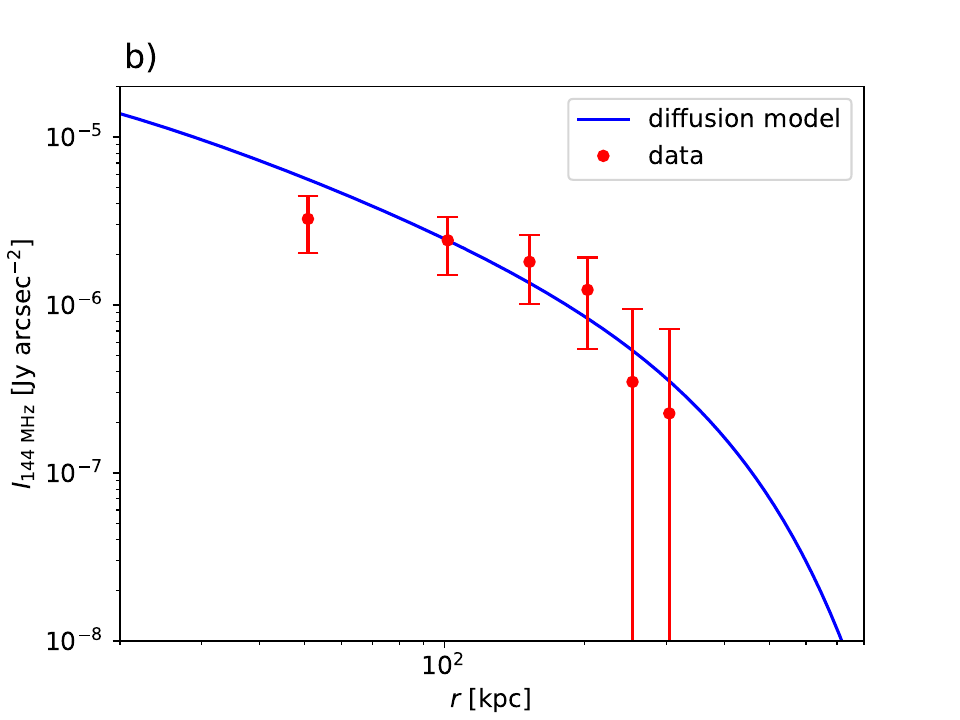}
\caption{Top (a): 144 MHz LOFAR 6\arcsec\ image of SPARCS1049+56, highlighting key components: S1 is the candidate radio mini-halo (Section~\ref{res:cl}). All other regions, shown in magenta, are excluded when computing the radio power of the mini-halo and when extracting the radio profile shown in the bottom panel. S2 is the star-forming region (Section~\ref{res:sf}); S3 and S4 are foreground galaxies, masked and discussed in Appendix~\ref{appB}; S5 and S6 are also masked; S7 is a possible contaminating AGN and also discussed in Appendix~\ref{appB}. Red dashed circles mark the radial bins used to extract the azimuthally averaged radio profile (bottom panel). Color bar units are Jy/beam. Bottom (b): Comparison of the azimuthally averaged radio profile at 144 MHz (red) and the radio emission predicted by the hadronic model (blue) for the radio mini-halo in SpARCS1049+56, assuming a diffusing CR distribution from the central AGN. We adopt a magnetic profile $B(r)=B_0(n_\mathrm{e}/n_0)^{1/3}$ with $B_0=15\,\mu$G.
}
\label{fig:hadronic_model}
\end{figure}

Two competing models explain the origin of mini-halos: the hadronic model, where CRs inelastically interact with ICM ions, producing pions that decay into secondary particles (e.g., electrons and positrons) emitting radio synchrotron radiation \citep{Pfrommer2004}. This requires relatively low energy density compared to the thermal pool \citep{Pfrommer2004_MEC}. Alternatively, a pre-existing aged CR electron population (at Lorentz factors of $\sim$100) may be re-accelerated via interactions with (turbulent) plasma waves \citep{Brunetti_2014}, which can be amplified by sloshing motions \citep[e.g.,][]{Giacintucci2024} or AGN jets interacting with the ICM \citep[][]{RL2020}.

Regardless of the mechanism, radio mini-halo emission is typically confined by cold fronts--sharp X-ray surface brightness discontinuities linked to sloshing motions--though some show faint extensions beyond the core \citep[e.g.,][]{Giacintucci2024}. Sloshing also amplifies and aligns magnetic fields along cold fronts via tangential shear flows \citep{Keshet2010} or magnetic draping \citep{Dursi2008}, preventing CRs from crossing fronts and constraining radio emission within these boundaries.

The origin of CR protons (hadronic model) or electrons (re-acceleration model) likely traces back to the central AGN, as inferred from correlations between mini-halo power and AGN feedback indicators such as AGN radio luminosity and X-ray cavity power \citet[][]{RL2020}. Supporting this, \citet[][]{Lusetti2024} recently identified extended radio emission around the mini-halo in Abell 1413. They argue that the mini-halo and its extended counterpart exhibit distinct properties, suggesting different mechanisms drive particle acceleration in these regions.

SpARCS1049+56 shows no evidence of recent AGN feedback from the BCG \citep[see][]{trudeau_multiwavelength_2019}. Deep \textit{Chandra} X-ray observations led \citet[][]{Hlavacek_Larrondo_2020} to propose that its intense $860\pm130$ M$_{\rm \odot}$ yr$^{-1}$ starburst results from catastrophic ICM cooling in the absence of AGN feedback. They suggested that the apparent cool core displacement from the BCG--likely due to an interaction offsetting the ICM and BCG, with the ICM sloshing back toward the new gravitational center--prevents AGN fueling, stalling feedback. As a result, the core cools catastrophically, sustaining the starburst. This implies that SpARCS1049+56 lacks AGN feedback but exhibits strong sloshing, and thus should not show turbulence induced by AGN feedback.

The discovery of a radio mini-halo at $z=1.709$ may not be entirely unexpected. The synchrotron emissivity scales as:
\begin{equation}
j_{\nu} \propto C_{e}  \frac{B^{1-\alpha}}{B^2 + B^2_{\text{CMB}}}\nu^{\alpha},
\end{equation}
where the normalization of the CR electron population, $C_e$, is given by
\begin{eqnarray}
    C_e&\propto\varepsilon_p\rho~~~
    \quad&\mbox{in the hadronic model and}\\
    C_e&\propto\eta_{\text{rel}} \frac{\displaystyle\rho v_t^3}{\displaystyle L_{\text{inj}}} 
    \quad&\mbox{in the re-acceleration model.}    
\end{eqnarray}
Here, $\varepsilon_p$ is the CR proton energy density \citep[hadronic model,][]{Pfrommer2004}, \( (\rho v_t^3)/L_{\text{inj}} \) is the turbulent energy flux \citep[re-acceleration model,][where \( \rho \) is the gas density, \( v_t \) the turbulent velocity and \( L_{\text{inj}} \) the injection scale]{DiGennaro2020} and \( \eta_{\text{rel}} \) is the fraction of turbulent energy flux reaccelerating seed relativistic electrons. \( B \) is the magnetic field strength, and \( B_{\text{CMB}} = 3.24\,(1 + z)^2 \, = 23.8\, \mu \text{G}\) is the cosmic microwave background (CMB) equivalent magnetic field strength at $z=1.709$. Hence, unless the magnetic field can be efficiently amplified to $\sim$$B_{\text{CMB}}$, the radio power is expected to be weaker due to increased inverse Compton losses via the CMB term. Interestingly, cosmological zoom simulations of massive galaxy clusters incorporating up-to-date galaxy formation physics predict a sufficiently fast magnetic dynamo in the proto-cluster region, saturating in the cluster core at $z\approx4$ with strengths exceeding 10~$\mu$G \citep{Tevlin_2024}. 

\begin{figure}
\centering
\includegraphics[width = 0.45\textwidth]{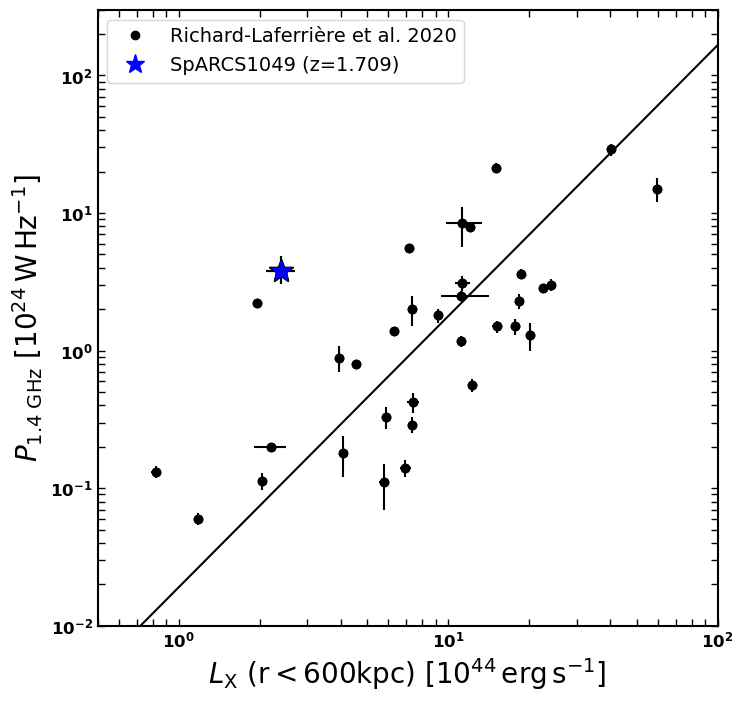}
\caption{Compilation of mini-halos from \citet{RL2020}, along with the best fit shown with the solid line, also from \citet{RL2020}. The mini-halo radio power ($P_{\rm 1.4\,GHz}$) is shown as a function of X-ray luminosity within a radius of 600 kpc ($L_{\rm X}$). SpARCS1049+56 is represented by the blue star. Here, we used the flux density at 144 MHz from the LOFAR 6\arcsec\/ image and extrapolated it to 1.4 GHz, while the X-ray luminosity was estimated using XMM-Newton observations.}
\label{fig:plot}
\end{figure}

For the re-acceleration model, following \citet[][]{DiGennaro2020} and their Equations 4–7, and assuming the same one-third mass ratio between the merging clusters, we find that the turbulent energy flux at $z=1.709$ could be 10 to 35 times higher than in the local Universe \citep[$z\sim0.2$;][]{DiGennaro2020}. Thus, with a $\mu$G-level magnetic field at $z=1.709$, comparable to those at $z\sim0.2$, a mini-halo could form within the scatter of Fig.~\ref{fig:plot} under the re-acceleration model. However, this would render hydrostatic mass estimates invalid due to the high turbulent pressure support.

For the hadronic model and to predict the synchrotron brightness profile of hadronically generated secondary electrons/positrons, we need to adopt a spatial distribution of CR protons and of the magnetic field. The spatial CR proton distribution is modeled with isotropic diffusion while the momentum distribution obeys a power-law, $f\propto p^{-2.3}$ and a low-momentum cutoff of $m_{p} c$. This choice implies a radio spectral index $\alpha=-1.15$ of the steady-state secondary electron population. The relatively flat observed synchrotron profile (Fig.~\ref{fig:hadronic_model}, panel b) requires a weak dependence of the magnetic field strength on gas density so that we adopt $B(r)=B_0(n_\mathrm{e}/n_0)^{\alpha_B}$, where $B_0 = 15\,\mu$G, $\alpha_B =1/3$ and $n_\mathrm{e}$ is the electron density as observed by Chandra \citep[][]{Hlavacek_Larrondo_2020}. This scaling can be realized by flux-freezing magnetic fields that are compressed along along the parallel and one perpendicular direction relative to the magnetic field or by taking a saturated small-scale dynamo with $\alpha_B=1/2$ \citep{Tevlin_2024} that experiences magnetic reconnection, implying that magnetic field lines can slip through plasma, rather than being compressed with it so that the density dependence of $B$ weakens. Isotropic compression and flux freezing gives $\alpha_B=2/3$ and would result in profiles that are too peaked.


In our scenario, CRs have been produced by the accreting central supermassive black hole (SMBH) during the proto-cluster assembly from $z\sim8$ to 4 \citep{Weinberger2017,Tevlin_2024}, which is centered on an age of the universe of $\sim$1 Gyr at $z\sim6$. We assume a rough SMBH mass of $\sim10^9~\mathrm{M}_\odot$ based on BCG scaling relations \citep{webb_extreme_2015}. If 5\% of its rest-mass energy is converted to CRs (i.e., $E_p = 9 \times 10^{61}$~erg) through acceleration in relativistic jets or quasar winds, and if these CR protons diffuse over $\sim$3 Gyr (until $z\approx 1.7$) via a combination of self-generated plasma waves (on small scales) and turbulent advection (on large scales) with an isotropic diffusion coefficient of $5\times10^{30}~\mathrm{cm}^2~\mathrm{s}^{-1}$ \citep{Pfrommer2003,Quataert2025}, then their distribution can extend to the edges of the candidate radio mini-halo.

Those CR protons generate secondary electrons via hadronic reactions that emit radio synchrotron emission at the observed level (see bottom panel of Fig.~\ref{fig:hadronic_model}, which reproduces the observed data well and also agrees with the innermost data point at the 2$\sigma$ level). The integrated CR-to-thermal energy ratio amounts to 0.07 within 200~kpc and decreases quickly with increasing radius due to the exponential cutoff. The high source redshift implies a considerable surface brightness dimming and K-correction by a factor of $(1+z)^{3-\alpha}\approx62.5$ (see Appendix~\ref{app:hadronic}), which is partially compensated for by accounting for a clumped distribution with spatial correlations among the gas density, CRs and magnetic field, which produce the observed patchy radio synchrotron signal (see top panel of Fig.~\ref{fig:hadronic_model}). We adopt a conservative choice for these spatial correlations that are manifested in the hadronic clumping factor $\mathcal{C}_\mathrm{2}=\langle n^2\rangle/\langle n\rangle^2=16$, which is calculated from our cosmological simulation of a cluster that is similar in mass to SpARCS1049+56 at $z=1.7$ (see Appendix~\ref{app:clumping} for more detail).
We verified that the hadronic and Alfv\'en cooling times in the bulk of the ICM are longer than the 3 Gyr (see Appendix~\ref{app:CR_cooling}) so that those CR protons do not appreciably lose their energy during the diffusion process.

These results suggest that, as in the re-acceleration model, a $10~\mu$G magnetic field at $z=1.709$ could produce a mini-halo within the scatter of Fig.~\ref{fig:plot}. Despite increased IC losses, an efficient magnetic dynamo could sustain low-redshift-like radio powers with moderate CR energy densities, easily injected by AGN jets \citep{Ehlert2018}. The detection of a mini-halo at $z=1.709$ may thus be less surprising, pointing to effective magnetic amplification during cluster formation.

\section{Concluding remarks} \label{sec:conc}

The discovery of a candidate radio mini-halo in a $z=1.709$ cluster of galaxies represents a significant step towards understanding diffuse radio emission in these structures. This detection was made possible by the remarkable sensitivity of LOFAR and the steep-spectrum nature of diffuse radio sources in galaxy clusters. 

Given the nature of this cluster—its high mass for its redshift, strong cool-core characteristics, and evidence of ICM sloshing (given the displaced cool core from the BCG)—it is perhaps not surprising to detect such a structure in this system. Importantly, our findings suggest that such structures may have formed very early in the lifetime of galaxy clusters. This implies that relativistic particles and strong magnetic fields were already in place as early as $z\sim1.7$ (i.e. when the Universe was $\sim3.8$ Gyrs old), and that either reacceleration processes or hadronic mechanisms were active during the epoch of cluster formation. Consequently, clusters of galaxies, including their progenitors (proto-clusters), may have been immersed in relativistic particles for most of their lifetimes. This environment may influence the galaxies residing within these clusters, potentially also shaping their evolution.
This study provides new insights into the evolution of large-scale structures and the role of diffuse radio emission over cosmic time. The advent of next-generation facilities, particularly high-sensitivity and high-resolution surveys with the SKA and ngVLA, will undoubtedly advance this field even further, enabling the exploration of these phenomena in unprecedented detail.

\section*{Acknowledgments}
JHL acknowledges funding support from the Canada Research Chairs Program, as well as the Natural Sciences and Engineering Research Council of Canada (NSERC) through the Discovery Grant and Accelerator Supplement programs. RT is grateful for support from the UKRI Future Leaders Fellowship (grant MR/T042842/1). CP thanks Ewald Puchwein for helpful discussions and acknowledges support by the European Research Council under ERC-AdG grant PICOGAL-101019746, by the German Science Foundation (DFG) through the Research Unit FOR-5195, and gratefully acknowledges the Gauss Centre for Supercomputing (GCS) for providing computing time on the GCS Supercomputer SuperMUC-NG at the Leibniz Supercomputing Centre (LRZ) in Garching, Germany, under project pn68cu. This work was supported by the STFC [grants ST/T000244/1, ST/V002406/1]. MLGM acknowledges financial support from NSERC via the Discovery grant program and the Canada Research Chair program. GDG acknowledges support from the ERC Consolidator Grant ULU 101086378.

The Université de Montréal recognizes that it is located on unceded (no treaty) Indigenous territory, and wishes to salute those who, since time immemorial, have been its traditional custodians. The University expresses its respect for the contribution of Indigenous peoples to the culture of societies here and around the world. The Université de Montréal is located where, long before French settlement, various Indigenous peoples interacted with one another. We wish to pay tribute to these Indigenous peoples, to their descendants, and to the spirit of fraternity that presided over the signing in 1701 of the Great Peace of Montr\'eal, a peace treaty founding lasting peaceful relations between France, its Indigenous allies and the Haudenosauni Confederacy (pronounced: O-di-no-sho-ni). The spirit of fraternity that inspired this treaty is a model for our academic community.

The Dunlap Institute is funded through an endowment established by the David Dunlap family and the University of Toronto.

This paper is based on data obtained with the LOFAR telescope (LOFAR-ERIC) under project code LT16\_005. LOFAR (van Haarlem et al. 2013) is the Low Frequency Array designed and constructed by ASTRON. It has observing, data processing, and data storage facilities in several countries, that are owned by various parties (each with their own funding sources), and that are collectively operated by the LOFAR European Research Infrastructure Consortium (LOFAR-ERIC) under a joint scientific policy. The LOFAR-ERIC resources have benefited from the following recent major funding sources: CNRS-INSU, Observatoire de Paris and Université d'Orléans, France; BMBF, MIWF-NRW, MPG, Germany; Science Foundation Ireland (SFI), Department of Business, Enterprise and Innovation (DBEI), Ireland; NWO, The Netherlands; The Science and Technology Facilities Council, UK; Ministry of Science and Higher Education, Poland.


\appendix

\section{LOFAR Observations: additional images}\label{appA}

\begin{figure*}
\centering
\includegraphics[width = 1\textwidth]{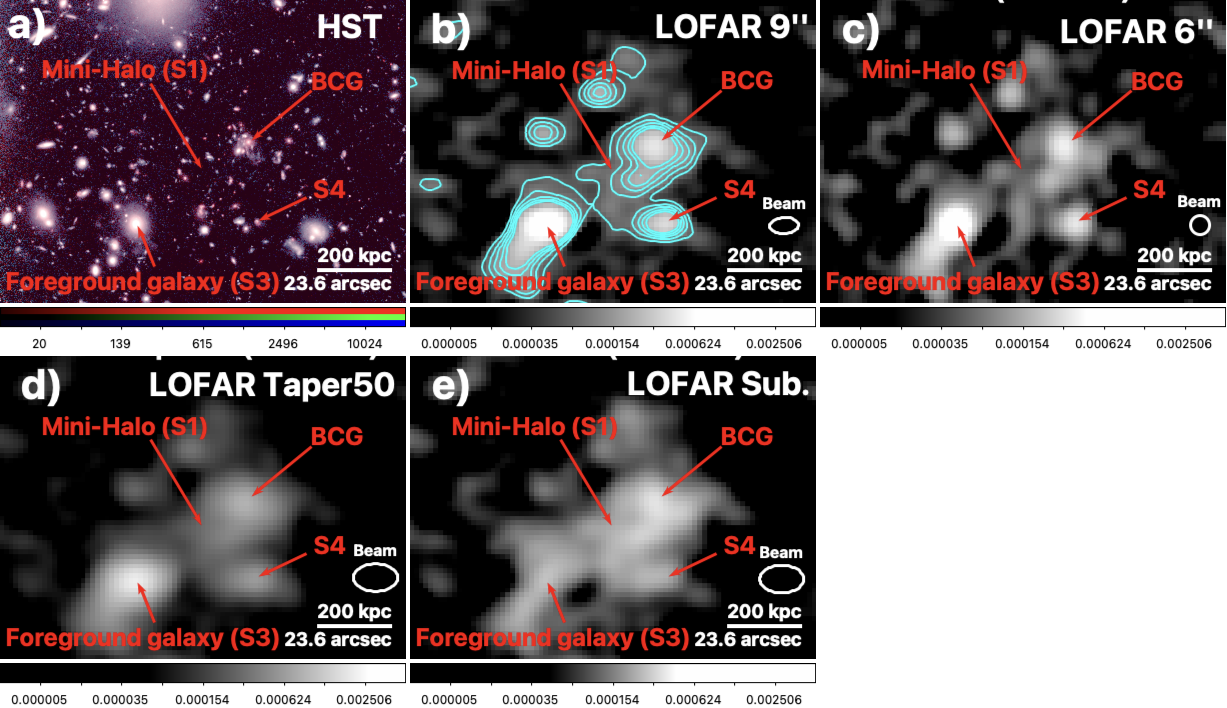}
\caption{Top-Left (a): Composite HST image (F105W in blue and green; F160W in red) of the SpARCS1049+56, with contours from the top-middke. Top-middle (b): 144 MHz LOFAR 9\arcsec\/ image with a beam of 8.9\arcsec$\times$5.9\arcsec. Contours are 2, 3, 4, 5 and 6$\sigma_{\rm rms}$ where $\sigma_{\rm rms}=35\mu$Jy/beam (also shown in panel a). Top-right (c): 144 MHz LOFAR 6\arcsec\/ image with a beam of 6.0\arcsec$\times$6.0\arcsec and $\sigma_{\rm rms}=33\mu$Jy beam$^{-1}$. Bottom-left (d): 144 MHz LOFAR image but with a $uv$ taper of 50 kpc. The resulting beam size is 13.8\arcsec$\times$8.6\arcsec\/ and $\sigma_{\rm rms}=53\mu$Jy beam$^{-1}$. Bottom-middle (e): attempted point source subtraction as detailed in \citet{Osinga_2021}. This attempt is only partially successful since the emission still shows peaks in the location of the compact sources seen in the 6\arcsec\/ image. Here, the resulting beam is 13.8\arcsec$\times$8.6\arcsec and $\sigma_{\rm rms}=43\mu$Jy beam$^{-1}$. This figure shows the presence of a diffuse extended emission in all radio images. We provide a zoom-in of the foreground galaxies labeled as S3 and S4 in Fig.~\ref{fig:supp2}. Color bars are included; radio image units are Jy/beam.}
\label{fig:supp1}
\end{figure*}

In Fig.~\ref{fig:supp1}, we present the re-processed LOFAR images from the Deep Fields. In the left panel, we display the re-processed image with a beam of 9\arcsec\ by 5\arcsec\ and $\sigma_{\rm rms}=35\mu$Jy beam$^{-1}$. The middle-left panel shows the image with a beam size of 6\arcsec\ by 6\arcsec\ and $\sigma_{\rm rms}=33\mu$Jy beam$^{-1}$. In the middle-right panel, we applied a $uv$ taper of 50 kpc to enhance the visibility of more extended features, resulting in a beam size of 13.8\arcsec\ by 8.6\arcsec\ and $\sigma_{\rm rms}=53\mu$Jy beam$^{-1}$. Although the diffuse emission is less spatially resolved, it remains detectable. Finally, the right panel shows an attempt to subtract point sources from the middle-right panel, as described in \cite{Osinga_2021}, with the resulting beam of 13.8\arcsec\ by 8.6\arcsec\ and $\sigma_{\rm rms}=43\mu$Jy beam$^{-1}$. The diffuse extended emission is detected in all panels, although the first two highlight the mini-halo the best due to their higher spatial resolution.

\section{Foreground galaxies}\label{appB}

\begin{figure*}
\centering
\includegraphics[width = 1\textwidth]{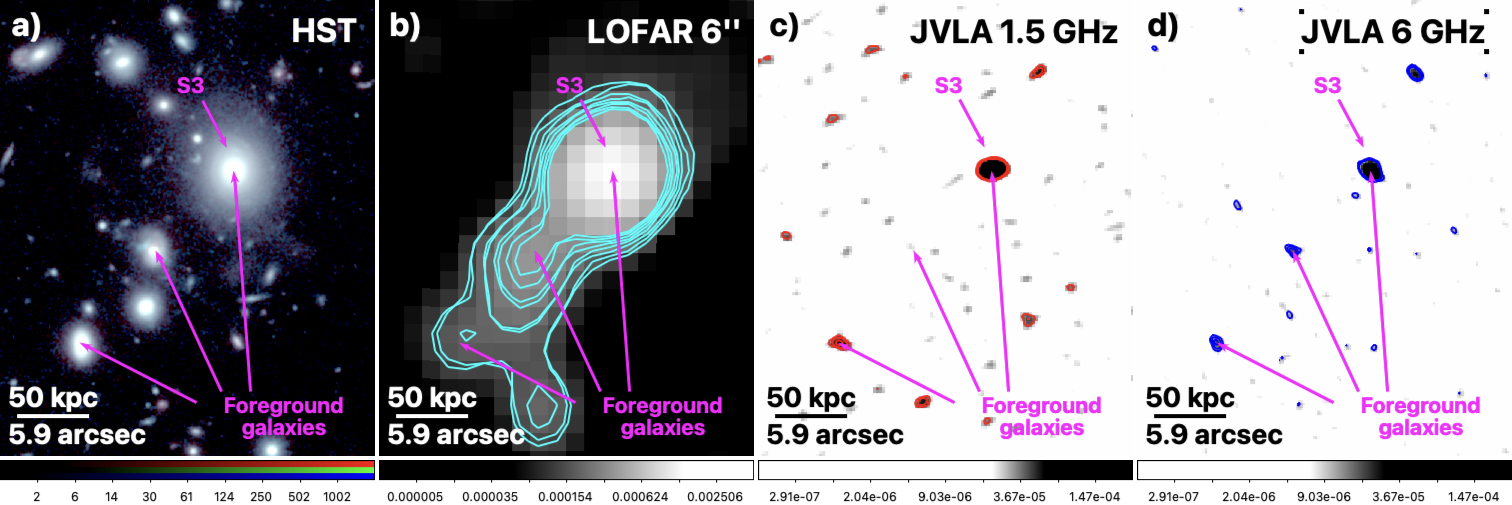}
\includegraphics[width = 1\textwidth]{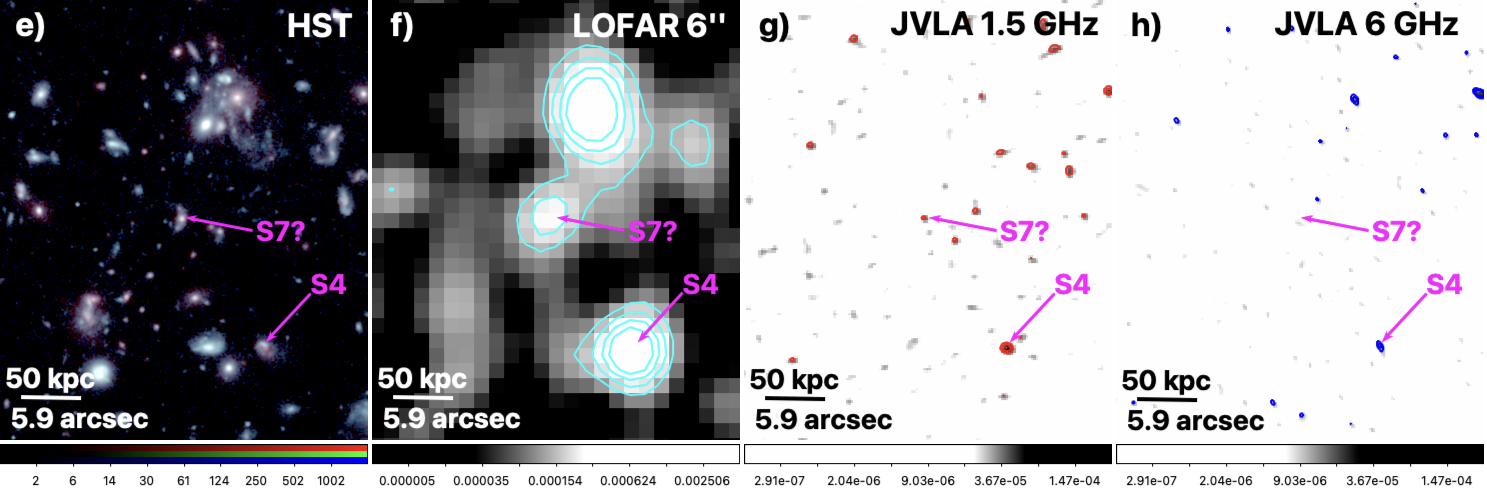}
\caption{Top: Zoom-in on the foreground $z\sim0.35$ galaxy located to the south-east of SPARCS1049+56 and labeled S3 in Fig.~\ref{fig:hadronic_model}. 
Left (a): Composite HST image (F105W in blue and green; F160W in red), highlighting foreground galaxies, including S3. Middle-left (b): 144 MHz LOFAR low-resolution image with a beam size of 6.0\arcsec\/ by 6.0\arcsec, where contours are 4, 6, 8 and 10$\sigma_{\rm rms}$ and $\sigma_{\rm rms}=33\mu$Jy/beam. 
Middle-right (c): JVLA 1.5 GHz image from \cite{trudeau_multiwavelength_2019}, with contours at 3, 4 and 5$\sigma_{\rm rms}$, where $\sigma_{\rm rms}=10.5\mu$Jy/beam. 
Right (d): JVLA 6 GHz image from \cite{trudeau_multiwavelength_2019}, with contours at 3, 4 and 5$\sigma_{\rm rms}$, where $\sigma_{\rm rms}=3\mu$Jy/beam. 
Although LOFAR looks like a one-sided jet, the combined JVLA and HST images show that the radio emission originates from multiple foreground galaxies and their associated AGN.
Bottom: Same, but zoomed-in on the source S4 and the source S7? from Fig.~\ref{fig:hadronic_model}. Color bars are included; radio image units are Jy/beam.}
\label{fig:supp2}
\end{figure*}

Figure~\ref{fig:largescale} highlights a foreground galaxy at $z \sim 0.35$, located southeast of the cluster, and identified as S3 in Fig.~\ref{fig:hadronic_model}.
This galaxy has a photometric redshift of $z = 0.346$, according to SDSS. In the top panels of Fig.~\ref{fig:supp2}, we present a zoom-in view of this galaxy, along with several other galaxies located to the south-east. All of the highlighted galaxies have redshifts between 0.4 and 0.6 according to SDSS, identifying them as foreground sources. Additionally, we show the radio emission associated with these galaxies. While the low-resolution 144 MHz LOFAR image suggests the presence of radio structure originating from the foreground $z\sim0.35$ galaxy resembling a one-sided radio jet, the higher-resolution JVLA 1.5 GHz and 6 GHz images, combined with HST data, reveal that this emission actually originates from foreground galaxies and their associated radio AGN. Therefore, the emission linked to these foreground galaxies is not a one-sided jet but rather a superposition of unrelated foreground sources. Consequently, the diffuse emission associated with SPARCS1049+56, located to the north-west of these galaxies, is distinct and, as argued in this paper, is consistent with a radio mini-halo. This result highlights the importance of high-resolution, multi-wavelength imaging in accurately disentangling complex radio structures, especially in crowded high redshift fields. Radio emission from these foreground sources is excluded when estimating the mini-halo's power and extracting the profile in panel b of Fig.~\ref{fig:hadronic_model}.

The bottom panels of Fig.~\ref{fig:supp2} show a zoom-in on source S4, located south of the BCG and west of S3 (see Fig.~\ref{fig:hadronic_model}). Detected at 1.5 and 6 GHz with the JVLA, S4 is consistent with an AGN. Its radio emission is excluded from both the mini-halo power estimate and the profile in panel b of Fig.~\ref{fig:hadronic_model}. We also mark a potential contaminant, labeled S7? in Fig.~\ref{fig:hadronic_model}. While its nature remains uncertain, S7? aligns with a galaxy (panel e) and a faint 1.5 GHz AGN (panel g) in Fig.~\ref{fig:supp2}, but is not detected in the LOFAR International data. Excluding S7? reduces the mini-halo power by just 5 per cent, and thus does not impact on our results.

\section{Modeling the radio emission}
\label{appC}
\subsection{Hadronic emission model}
\label{app:hadronic}

Inelastic collisions between CR nuclei and the thermal ICM generate secondary radio synchrotron-emitting CR electrons. We assume that the CR proton distribution follows a power-law in momentum\footnote{Note that we redefine the sign of the spectral indices in this appendix for convenience, so that the synchrotron spectral index in the main part of the paper, $\alpha$, can be obtained via the transformation $\alpha = -\alpha_\nu$.} at position $\x$, $f_p(p,\x)=C_{p}^{}(\x)p_{p}^{-\alpha_p}\Theta_\mathrm{H}(p_p-q)$, where $p_p=\tilde{p}_p/(m_{p}c)$ is the dimensionless proton momentum (and $m_{p}$ and $c$ denote the proton rest mass and light speed, respectively), $\Theta_\mathrm{H}$ is the Heaviside function, $q$ is the dimensionless low-momentum cutoff and $C_p$ is implicitly given in terms of the CR energy density \citep{Pfrommer2003}
\begin{eqnarray}
\label{eq:eps_p}
\varepsilon_p(\x) &=& 
\frac{C_p(\x)\,m_p \,c^2}{2\,(\alpha_p-1)}\,
\left[\mathcal{B}_{\frac{1}{1+q^2}}\left(\frac{\alpha_p-2}{2},
\frac{3-\alpha_p}{2}\right) + 2\,q^{1-\alpha_p} 
\left( \sqrt{1+q^2} - 1 \right)\right].
\end{eqnarray}
Here $\mathcal{B}_x(a,b)$ denotes the incomplete beta-function. We adopt a proton spectral index $\alpha_p=2.3$ and a low-momentum cutoff $q=1$ (which is generated as a result of significant Coulomb interactions of CR protons and the ICM over $\sim$3~Gyr, see \citealt{Ensslin2007}). As described in the main text, in our model the spatial CR proton distribution at $z=1.709$ emerges from a \textit{CR diffusion} process, where the majority of CRs has been produced close to the cluster center during the time of proto-cluster assembly at around $z\sim6$ (by either accretion processes or outflows from the central AGN and the corresponding initial starburst). The solution of the diffusion equation for a Dirac delta function source at the cluster center in both time and space yields \citep{Pfrommer2003}
\begin{eqnarray}
\label{eq:CR_diff}
\varepsilon_p(r,t) &=& 
\frac{E_p}{(4\pi\kappa t)^{3/2}}\,
\exp\left(-\frac{r^2}{4\kappa t}\right),
\end{eqnarray}
where $E_p$ is the CR proton energy injected at the center, $\kappa$ is the isotropic CR diffusion coefficient, and $r$ and $t$ are the radial coordinate and time elapsed since the injection event. The injected secondary electrons cool radiatively via inverse Compton and synchrotron processes.  At high electron momenta, $p_e=\tilde{p}_e/(m_{e}c)>\mbox{GeV}/c$ (where $m_e$ denotes the electron rest mass), those processes balance each other, yielding an equilibrium distribution of secondary CR electrons \citep{Pfrommer2008}\footnote{We neglect the (minor) reduction in the electron distribution within the transition regime compared to its steady-state value, as this effect can be compensated by adopting a steeper injection spectral index \citep{Pfrommer2017}.}
\begin{eqnarray}
\label{eq:fe_hadr}
f_e(p_e,\x)\,dp_e&=&C_{e}(\x)p_e^{-\alpha_e}\,dp_e,\\
C_e(\x)&=&\frac{16^{2-\alpha_e}\sigma_{pp}\,n_{n}(\x)C_p(\x)\,m_{e}c^2}
     {(\alpha_e-2)\,\sigma_\mathrm{T}\,[\varepsilon_B(\x)+\varepsilon_\mathrm{CMB}]}\left(\frac{m_p}{m_e}\right)^{\alpha_e-2},
\label{eq:Ce_hadr}
\end{eqnarray}
where $\alpha_e=\alpha_p+1$ is the cooled electron spectral index, $n_{n}=n_\mathrm{H}+4n_\mathrm{He}=\rho/m_p$ is the target density of nucleons (neglecting metals), $\sigma_\mathrm{T}$ is the Thomson cross section, $\varepsilon_B$ and $\varepsilon_\mathrm{CMB}$ are the energy densities of the magnetic field and the CMB, respectively.  The cross section $\sigma_{pp}$ parameterizes all processes at the pion-production threshold in terms of the proton spectral index \citep{Pfrommer2004},
\begin{equation}
\label{eq:sigmapp}
\sigma_{pp}\simeq32\times\left(0.96+\mathrm{e}^{4.4\,-\,2.4\,\alpha_p}\right)~\mbox{mbarn}. 
\end{equation}

The synchrotron intensity depends on the strength of the transverse (with respect to the line-of-sight) component of the magnetic field, and the spatial and spectral distribution of CR electrons. Those electrons are assumed to be isotropically distributed in pitch angle (i.e., the angle between the electron's momentum vector and the magnetic field). The intensity of the emissivity, $j_\nu\equiv{d}E/(dt\,d\nu\,d^3x\,d\Omega)$, for such a power-law
momentum distribution of electrons is given by \citep{Rybicki1979}:
\begin{eqnarray}
\label{eq:jnu}
j_\nu(\x)&=&\frac{\sqrt{3\pi}e^3}{16\pi m_{e}c^2}\,\left(\frac{4\pi{\nu}m_{e}c}{3e}\right)^{-\alpha_\nu}B(\x)^{(\alpha_e+1)/2}C_e(\x)
\nonumber\\
&&\times
\frac{2^{(\alpha_e-3)/2}}{3}\left(\frac{3\alpha_e+7}{\alpha_e+1}\right)
\Gamma\left(\frac{3\alpha_e-1}{12}\right)\Gamma\left(\frac{3\alpha_e+7}{12}\right)
\Gamma\left(\frac{\alpha_e+5}{4}\right)\Gamma\left(\frac{\alpha_e+7}{4}\right)^{-1},
\end{eqnarray}
where $\alpha_\nu=(\alpha_e-1)/2=\alpha_p/2$ denotes the radio spectral index and $\Gamma(a)$ is the gamma function \citep{Abramowitz1965}. The radio surface brightness $I_\nu$ is obtained by integrating $j_{\nu}(\x)$ along the line-of-sight $s$ and is a function of observational frequency $\nu$ and position on the sky, $\x_\perp$, and reads (in units of erg~s$^{-1}$~Hz$^{-1}$~cm$^{-2}$~sterad$^{-1}$)
\begin{equation}
  \label{eq:Inu}
  I_\nu(\x_\perp)=\int_0^{\infty} j_\nu(\x_\perp,s)d s.
\end{equation}

Because the candidate radio mini halo is associated with a cluster at a cosmological distance, we need to take into account the effects of redshift dimming and K-correction. The rest frame values of frequency and emitted energy are redshifted as photons propagate through the universe so that we observe $d\nu=d\nu_\mathrm{rest}/(1+z)$ and $dE=dE_\mathrm{rest}/(1+z)$, while the time interval of the emitted radiation is dilated, $dt=(1+z)dt_\mathrm{rest}$. Note that $dE_\mathrm{rest}$ is the equivalent isotropically emitted energy during the time interval  $dt_\mathrm{rest}$. The differential solid angle of the source as seen by the observer is $d\Omega = d A_\mathrm{source}/D_\mathrm{ang}^2$ and the density of photons is diluted as they propagate radially away from the source on an expanding sphere so that the fraction of radiation received per unit collection area is $dA/(4\pi D_\mathrm{com}^2)$, where the comoving distance between the source and the observer is $D_\mathrm{com}=(1+z) D_\mathrm{ang}$ in a flat universe. The observed surface brightness is thus given by 
\begin{equation}
  \label{eq:Inu_z}
  I_\nu\equiv\frac{dE}{dt\,d\nu\,dA\,d\Omega}\frac{dA}{4\pi D_\mathrm{com}^2}
  =I_{\nu_\mathrm{rest}}\,\frac{1}{(1+z)^3}
  =I_0\nu^{-\alpha_\nu}\,\frac{1}{(1+z)^{\alpha_\nu + 3}},
\end{equation}
where $I_{\nu_\mathrm{rest}}=dE_\mathrm{rest}/(dt_\mathrm{rest}\,d\nu_\mathrm{rest}\, dA\,d\Omega_\mathrm{close})\,\times dA/(4\pi D_\mathrm{phy}^2)$ is the surface brightness at a distance close to the object (which implies that cosmological effects are negligible so that  $d\Omega_\mathrm{close} = d A_\mathrm{source}/D_\mathrm{phy}^2$ where $D_\mathrm{phy}=D_\mathrm{ang}$) and we assume a power-law radio spectrum in the last step, $I_{\nu_\mathrm{rest}}=I_0\nu_\mathrm{rest}^{-\alpha_\nu}$. The well-known result of redshift dimming of the bolometric surface brightness \citep{Tolman1934} is obtained by means of integrating over the frequency range, to arrive at 
\begin{equation}
  \label{eq:Inu_bol}
  I = \int_0^\infty I_\nu d\nu 
  = \frac{1}{(1+z)^4}\, \int_0^\infty I_{\nu_\mathrm{rest}} d\nu_\mathrm{rest}
  = \frac{1}{(1+z)^4}\, I_\mathrm{rest}.
\end{equation}
The observed specific flux is related to the specific power (or luminosity) via
\begin{equation}
  \label{eq:Snu}
  S_\nu = \int_{\partial\Omega} I_\nu d\Omega = \frac{P_{\nu_\mathrm{rest}}}{4\pi D_\mathrm{lum}^2} (1+z)
  = \frac{P_{\nu}}{4\pi D_\mathrm{lum}^2} (1+z)^{1-\alpha_\nu},
\end{equation}
where $P_{\nu_\mathrm{rest}}=4\pi\int_{\partial A_\mathrm{source}} I_{\nu_\mathrm{rest}}d A_\mathrm{source}$ is the rest-frame specific luminosity, $P_{\nu}=P_0\nu^{-\alpha_\nu}$ is the rest-frame  specific luminosity observed at frequency $\nu$, and the luminosity distance between the observer at $z=0$ and the source is related to the angular diameter distance via $D_\mathrm{lum}=(1+z)^2 D_\mathrm{ang}$.

\subsection{Modeling the effect of a clumped ICM}
\label{app:clumping}

Modeling hadronic interactions in SpARCS1049+56 requires the knowledge of the ICM. We adopt the electron temperature of $5.7$~keV and the electron density profile, $n_e = n_0 (1 + r^2/r_c^2)^{-3/2\,\beta}$, where $n_0=6.602\times 10^{-2}~\mathrm{cm}^{-3}$, $r_c=10.14$~kpc and $\beta=0.4$, both inferred from X-ray observations \citep{Hlavacek_Larrondo_2020}. The thermal energy density is given by $\varepsilon_\mathrm{th}=3/2\times n k_\mathrm{B} T = 3/2\times d_e n_e k_\mathrm{B} T_e$, where $d_e=(3+5 X_\mathrm{H})/(2X_\mathrm{H}+2)\approx1.93$ is the number of particles per electron in a fully ionized primordial plasma with a hydrogen mass fraction of $X_\mathrm{H}=0.76$ and we assume $T=T_e$.

\begin{figure}
\centering
\includegraphics[width = 0.5\textwidth]{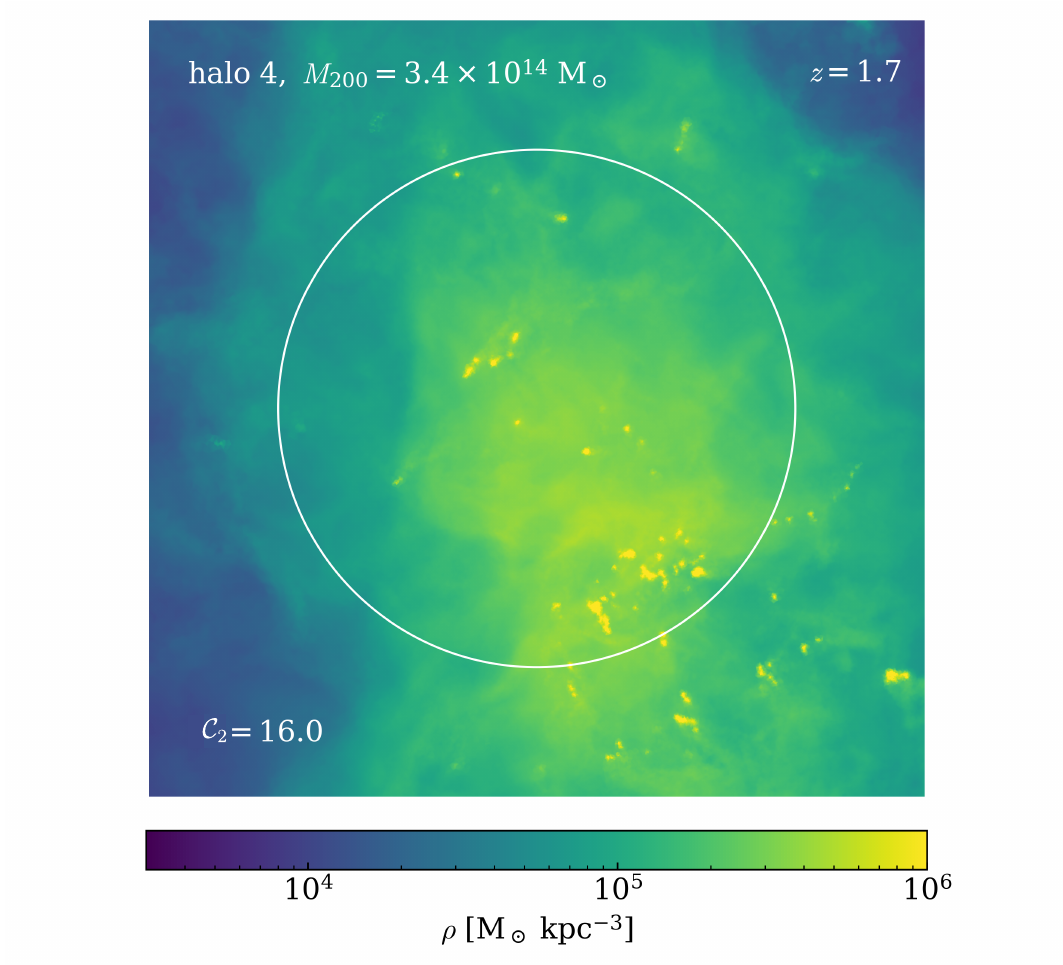}
\caption{Density distribution of a simulated galaxy cluster of mass $M_{200}=3.4\times10^{14}\,\mathrm{M}_\odot$ at redshift $z=1.7$ \citep{Tevlin_2024}. Shown is a projection with side length 600 kpc and projection depth 100 kpc. We obtain a clumping factor $\mathcal{C}_2=16$ within a radius of 200 kpc (indicated by the white circle), which is close to the maximum extend of the candidate radio mini-halo.}
\label{fig:clumping}
\end{figure}

Cosmological simulations show a clumped ICM at high redshift with spatial correlations among the gas density, magnetic fields, and CRs so that the resulting hadronically generated radio synchrotron emission should also be patchy. At high redshift, where inverse Compton cooling dominates the losses of high-energy electrons, the hadronic radio emission scales as $j_\nu\propto n C_p B^{1+\alpha_\nu}$ (as we can neglect the $B^2$ term in the denominator of Eq.~\ref{eq:Ce_hadr}). We assume a weak correlation of the magnetic field with gas density, $B\propto n^{1/3}$ (see Section~\ref{sec:astrophysics}) and that a combination of CR advection in the turbulent ICM and self-generated CR diffusion sets the correlation with the gas density. While turbulent advection yields $C_p\propto n^{(\alpha_p+2)/3}\approx n^{1.4}$ for $\alpha_p=2.3$, we obtain $C_p=\mathrm{constant}$ in the limiting regime of pure CR streaming and diffusion \citep{Ensslin2011}. While CR streaming and diffusion at self-generated waves should be dominant at small scales, turbulent advection will take over on larger scales. We thus adopt a conservative estimate of $C_p\propto n^{0.3}$, which enables us to define the hadronic clumping factor:
\begin{equation}
  \label{eq:clumping}
  \mathcal{C}_\mathrm{had} \equiv \frac{\langle n C_p B^{1+\alpha_\nu}\rangle}{\langle n \rangle \langle C_p\rangle \langle B^{1+\alpha_\nu}\rangle}
  \approx \frac{\langle n^2 \rangle}{\langle n \rangle^2} \equiv \mathcal{C}_2,
\end{equation}
where the brackets indicate volume averages. This clumping effect biases the expected radio intensity high in relation to a smooth distribution because of correlations in gas density, CRs and magnetic fields. We find a hadronic clumping factor of $\mathcal{C}_\mathrm{2} =16$ within 200~kpc in our cosmological simulation of a cluster that is comparable in mass to SpARCS1049+56 at $z=1.7$ \citep{Tevlin_2024}; see Fig.~\ref{fig:clumping} for a visual impression of the clumped density distribution. While projection effects partially smooth the resulting surface brightness, the largest excursions in the $j_\nu$ distribution for such large clumping factors should survive projection effects, giving rise to a patchy radio synchrotron emission map. This emphasizes the need to conduct cosmological magnetohydrodynamic simulations with anisotropically diffusing CRs that have been accelerated by accreting AGNs to better quantify the discussed effects. Note that hadronic clumping partially compensates cosmological redshift dimming.

\subsection{Cosmic ray energetics and cooling timescales}
\label{app:CR_cooling}

Mass accretion onto a SMBH releases gravitational binding energy
\begin{equation}
  \label{eq:E_AGN}
  E_\mathrm{SMBH} = \dot{M} c^2 \Delta t\approx 2\times10^{63}~\mathrm{erg}\,\left(\frac{M_\mathrm{SMBH}}{10^9\,\mathrm{M}_\odot}\right),
\end{equation}
where $\dot{M}$ is the mass accretion rate over a time interval $\Delta t$. Assuming a typical radiative efficiency of 0.1 and an effective CR acceleration efficiency of 0.5 implies an energy fraction of 0.05 that is channeled into CRs. This corresponds to the required CR energy in the CR diffusion model to explain the observed radio emission for a combination of diffusion time of 3~Gyr and CR diffusion coefficient of $5\times10^{30}~\mathrm{cm}^2\,\mathrm{s}^{-1}$ (see Section~\ref{sec:astrophysics}).

When considering CR diffusion for an extended period of $t_\mathrm{diff}\approx3$~Gyr, it is critical to confirm that every cooling time scale of CR protons above GeV energies is larger than the diffusion time. The two relevant cooling processes are hadronic CR interactions with the ambient gas as well as CR streaming losses at self-generated plasma waves. The hadronic loss time is given by \citep{Ruszkowski_2023}
\begin{equation}
  \label{eq:t_pp}
  t_{pp}=\frac{1}{0.5 \sigma_{pp} n_\mathrm{tar} c} \approx 6.6~\mathrm{Gyr} > t_\mathrm{diff},
\end{equation}
where we adopted a typical target gas density $\rho/m_p \approx 10^{-2}~\mathrm{cm}^{-3}$. The time scale associated with the excitation of resonant Alfv\'en waves $t_\mathrm{A}$ equals the CR loss time scale in the steady-state limit of streaming CRs and is given by \citep{Ruszkowski_2023}
\begin{equation}
\label{eq:t_A}
\dot\varepsilon_p =
-\left|\bm{v}_\mathrm{A}\bm\cdot\bm{\nabla}P_p\right|
\quad\Rightarrow\quad
t_\mathrm{A} = \frac{\varepsilon_p}{\left|\dot\varepsilon_p\right|}\approx4.2~\mathrm{Gyr}> t_\mathrm{diff},
\end{equation}
where we assume that CRs stream at velocities of order the Alfv\'en speed ${v}_\mathrm{A}={B}/\sqrt{4\pi\rho}\approx 28~\mathrm{km~s}^{-1}$ in a medium of mean density of $\rho/(\mu m_p) \approx 10^{-2}~\mathrm{cm}^{-3}$ and a volume-averaged magnetic field of 1~$\mu$G (with a mean molecular weight of $\mu=0.6$ for fully ionized gas). For our estimates, we adopt $P_p=\varepsilon_p/3$ and a CR gradient length of $L_p=\varepsilon_p/|\bm{\nabla}\varepsilon_p|\approx40$~kpc. This demonstrates that the proposed CR diffusion model is plausible based on morphological, energetic, and timescale considerations, and it motivates a more in-depth study using cosmological simulations.

\bibliography{sample631}{}

\begin{thebibliography}{}
\expandafter\ifx\csname natexlab\endcsname\relax\def\natexlab#1{#1}\fi
\providecommand{\url}[1]{\href{#1}{#1}}
\providecommand{\dodoi}[1]{doi:~\href{http://doi.org/#1}{\nolinkurl{#1}}}
\providecommand{\doeprint}[1]{\href{http://ascl.net/#1}{\nolinkurl{http://ascl.net/#1}}}
\providecommand{\doarXiv}[1]{\href{https://arxiv.org/abs/#1}{\nolinkurl{https://arxiv.org/abs/#1}}}

\bibitem[{{Abramowitz} \& {Stegun}(1965)}]{Abramowitz1965}
{Abramowitz}, M., \& {Stegun}, I.~A. 1965, {Handbook of mathematical functions with formulas, graphs, and mathematical tables}

\bibitem[{B{\'e}gin {et~al.}(2023)B{\'e}gin, Hlavacek-Larrondo, Rhea, Gendron-Marsolais, McNamara, van Weeren, Richard-Laferri{\`e}re, Guit{\'e}, Prasow-{\'E}mond, \& Haggard}]{begin2023extended}
B{\'e}gin, T., Hlavacek-Larrondo, J., Rhea, C., {et~al.} 2023, Monthly Notices of the Royal Astronomical Society, 519, 767

\bibitem[{{Biava} {et~al.}(2024){Biava}, {Bonafede}, {Gastaldello}, {Botteon}, {Brienza}, {Shimwell}, {Brunetti}, {Bruno}, {Rajpurohit}, {Riseley}, {van Weeren}, {Rossetti}, {Cassano}, {De Gasperin}, {Drabent}, {Rottgering}, {Edge}, \& {Tasse}}]{Biava2024}
{Biava}, N., {Bonafede}, A., {Gastaldello}, F., {et~al.} 2024, \aap, 686, A82, \dodoi{10.1051/0004-6361/202348045}

\bibitem[{{Bonafede} {et~al.}(2023){Bonafede}, {Gitti}, {La Bella}, {Biava}, {Ubertosi}, {Brunetti}, {Lusetti}, {Brienza}, {Riseley}, {Stuardi}, {Botteon}, {Ignesti}, {R{\"o}ttgering}, \& {van Weeren}}]{Bonafede2023}
{Bonafede}, A., {Gitti}, M., {La Bella}, N., {et~al.} 2023, \aap, 680, A5, \dodoi{10.1051/0004-6361/202347567}

\bibitem[{Boxelaar {et~al.}(2021)Boxelaar, {van Weeren}, \& Botteon}]{boxelaar21}
Boxelaar, J., {van Weeren}, R., \& Botteon, A. 2021, Astronomy and Computing, 35, 100464, \dodoi{https://doi.org/10.1016/j.ascom.2021.100464}

\bibitem[{{Bravi} {et~al.}(2016){Bravi}, {Gitti}, \& {Brunetti}}]{Bravi2016}
{Bravi}, L., {Gitti}, M., \& {Brunetti}, G. 2016, \mnras, 455, L41, \dodoi{10.1093/mnrasl/slv137}

\bibitem[{{Brunetti} \& {Jones}(2014)}]{Brunetti_2014}
{Brunetti}, G., \& {Jones}, T.~W. 2014, International Journal of Modern Physics D, 23, 1430007, \dodoi{10.1142/S0218271814300079}

\bibitem[{Cassano {et~al.}(2006)Cassano, Brunetti, \& Setti}]{Cassano_2006}
Cassano, R., Brunetti, G., \& Setti, G. 2006, Monthly Notices of the Royal Astronomical Society, 369, 1577, \dodoi{10.1111/j.1365-2966.2006.10423.x}

\bibitem[{Cassano {et~al.}(2019)Cassano, Botteon, Di~Gennaro, Brunetti, Sereno, Shimwell, Van~Weeren, Br{\"u}ggen, Gastaldello, Izzo, {et~al.}}]{cassano2019lofar}
Cassano, R., Botteon, A., Di~Gennaro, G., {et~al.} 2019, The Astrophysical Journal Letters, 881, L18

\bibitem[{{Cassano} {et~al.}(2023){Cassano}, {Cuciti}, {Brunetti}, {Botteon}, {Rossetti}, {Bruno}, {Simionescu}, {Gastaldello}, {van Weeren}, {Br{\"u}ggen}, {Dallacasa}, {Zhang}, {Akamatsu}, {Bonafede}, {Di Gennaro}, {Shimwell}, {de Gasperin}, {R{\"o}ttgering}, \& {Jones}}]{Cassano2023}
{Cassano}, R., {Cuciti}, V., {Brunetti}, G., {et~al.} 2023, \aap, 672, A43, \dodoi{10.1051/0004-6361/202244876}

\bibitem[{{de Gasperin} {et~al.}(2019){de Gasperin}, {Dijkema}, {Drabent}, {Mevius}, {Rafferty}, {van Weeren}, {Br{\"u}ggen}, {Callingham}, {Emig}, {Heald}, {Intema}, {Morabito}, {Offringa}, {Oonk}, {Orr{\`u}}, {R{\"o}ttgering}, {Sabater}, {Shimwell}, {Shulevski}, \& {Williams}}]{deGasperin2019}
{de Gasperin}, F., {Dijkema}, T.~J., {Drabent}, A., {et~al.} 2019, \aap, 622, A5, \dodoi{10.1051/0004-6361/201833867}

\bibitem[{{Di Gennaro} {et~al.}(2021{\natexlab{a}}){Di Gennaro}, {van Weeren}, {Cassano}, {Brunetti}, {Br{\"u}ggen}, {Hoeft}, {Osinga}, {Botteon}, {Cuciti}, {de Gasperin}, {R{\"o}ttgering}, \& {Tasse}}]{DiGennaro2021}
{Di Gennaro}, G., {van Weeren}, R.~J., {Cassano}, R., {et~al.} 2021{\natexlab{a}}, \aap, 654, A166, \dodoi{10.1051/0004-6361/202141510}

\bibitem[{{Di Gennaro} {et~al.}(2021{\natexlab{b}}){Di Gennaro}, {van Weeren}, {Cassano}, {Brunetti}, {Br{\"u}ggen}, {Hoeft}, {Osinga}, {Botteon}, {Cuciti}, {de Gasperin}, {R{\"o}ttgering}, \& {Tasse}}]{Di_Gennaro_2021}
---. 2021{\natexlab{b}}, \aap, 654, A166, \dodoi{10.1051/0004-6361/202141510}

\bibitem[{{Di Gennaro} {et~al.}(2021{\natexlab{c}}){Di Gennaro}, {van Weeren}, {Brunetti}, {Cassano}, {Br{\"u}ggen}, {Hoeft}, {Shimwell}, {R{\"o}ttgering}, {Bonafede}, {Botteon}, {Cuciti}, {Dallacasa}, {de Gasperin}, {Dom{\'\i}nguez-Fern{\'a}ndez}, {En{\ss}lin}, {Gastaldello}, {Mandal}, {Rossetti}, \& {Simionescu}}]{DiGennaro2020}
{Di Gennaro}, G., {van Weeren}, R.~J., {Brunetti}, G., {et~al.} 2021{\natexlab{c}}, Nature Astronomy, 5, 268, \dodoi{10.1038/s41550-020-01244-5}

\bibitem[{{Di Gennaro} {et~al.}(2023){Di Gennaro}, {Br{\"u}ggen}, {van Weeren}, {Simionescu}, {Brunetti}, {Cassano}, {Forman}, {Hoeft}, {Ignesti}, {R{\"o}ttgering}, \& {Shimwell}}]{DiGennaro2023}
{Di Gennaro}, G., {Br{\"u}ggen}, M., {van Weeren}, R.~J., {et~al.} 2023, \aap, 675, A51, \dodoi{10.1051/0004-6361/202345905}

\bibitem[{{Duncan} {et~al.}(2019){Duncan}, {Sabater}, {R{\"o}ttgering}, {Jarvis}, {Smith}, {Best}, {Callingham}, {Cochrane}, {Croston}, {Hardcastle}, {Mingo}, {Morabito}, {Nisbet}, {Prandoni}, {Shimwell}, {Tasse}, {White}, {Williams}, {Alegre}, {Chy{\.z}y}, {G{\"u}rkan}, {Hoeft}, {Kondapally}, {Mechev}, {Miley}, {Schwarz}, \& {van Weeren}}]{Duncan2021}
{Duncan}, K.~J., {Sabater}, J., {R{\"o}ttgering}, H.~J.~A., {et~al.} 2019, \aap, 622, A3, \dodoi{10.1051/0004-6361/201833562}

\bibitem[{Dursi \& Pfrommer(2008)}]{Dursi2008}
Dursi, L.~J., \& Pfrommer, C. 2008, \apj, 677, 993, \dodoi{10.1086/529371}

\bibitem[{{Ehlert} {et~al.}(2018){Ehlert}, {Weinberger}, {Pfrommer}, {Pakmor}, \& {Springel}}]{Ehlert2018}
{Ehlert}, K., {Weinberger}, R., {Pfrommer}, C., {Pakmor}, R., \& {Springel}, V. 2018, \mnras, 481, 2878, \dodoi{10.1093/mnras/sty2397}

\bibitem[{{En{\ss}lin} {et~al.}(2011){En{\ss}lin}, {Pfrommer}, {Miniati}, \& {Subramanian}}]{Ensslin2011}
{En{\ss}lin}, T., {Pfrommer}, C., {Miniati}, F., \& {Subramanian}, K. 2011, \aap, 527, A99, \dodoi{10.1051/0004-6361/201015652}

\bibitem[{{En{\ss}lin} {et~al.}(2007){En{\ss}lin}, {Pfrommer}, {Springel}, \& {Jubelgas}}]{Ensslin2007}
{En{\ss}lin}, T.~A., {Pfrommer}, C., {Springel}, V., \& {Jubelgas}, M. 2007, \aap, 473, 41, \dodoi{10.1051/0004-6361:20065294}

\bibitem[{{En{\ss}lin} \& {R{\"o}ttgering}(2002)}]{En_lin_2002}
{En{\ss}lin}, T.~A., \& {R{\"o}ttgering}, H. 2002, \aap, 396, 83, \dodoi{10.1051/0004-6361:20021382}

\bibitem[{Finner {et~al.}(2020)Finner, Jee, Webb, Wilson, Perlmutter, Muzzin, \& Hlavacek-Larrondo}]{Finner_2020}
Finner, K., Jee, M.~J., Webb, T., {et~al.} 2020, The Astrophysical Journal, 893, 10, \dodoi{10.3847/1538-4357/ab7bdb}

\bibitem[{Gendron-Marsolais {et~al.}(2017)Gendron-Marsolais, Hlavacek-Larrondo, van Weeren, Clarke, Fabian, Intema, Taylor, Blundell, \& Sanders}]{gendron2017deep}
Gendron-Marsolais, M., Hlavacek-Larrondo, J., van Weeren, R., {et~al.} 2017, Monthly Notices of the Royal Astronomical Society, 469, 3872

\bibitem[{{Gendron-Marsolais} {et~al.}(2020){Gendron-Marsolais}, {Hlavacek-Larrondo}, {van Weeren}, {Rudnick}, {Clarke}, {Sebastian}, {Mroczkowski}, {Fabian}, {Blundell}, {Sheldahl}, {Nyland}, {Sanders}, {Peters}, \& {Intema}}]{GM2021}
{Gendron-Marsolais}, M., {Hlavacek-Larrondo}, J., {van Weeren}, R.~J., {et~al.} 2020, \mnras, 499, 5791, \dodoi{10.1093/mnras/staa2003}

\bibitem[{Gennaro {et~al.}(2025)Gennaro, Brüggen, Moravec, Mascolo, van Weeren, Brunetti, Cassano, Botteon, Churazov, Khabibullin, Lyskova, de~Gasperin, Hardcastle, Röttgering, Shimwell, Sunyaev, \& Stanford}]{DiGennaro2025}
Gennaro, G.~D., Brüggen, M., Moravec, E., {et~al.} 2025, Limits and challenges of the detection of cluster-scale diffuse radio emission at high redshift: The Massive and Distant Clusters of WISE Survey (MaDCoWS) in LoTSS-DR2.
\newblock \doarXiv{2502.19273}

\bibitem[{Giacintucci {et~al.}(2019)Giacintucci, Markevitch, Cassano, Venturi, Clarke, Kale, \& Cuciti}]{giacintucci2019expanding}
Giacintucci, S., Markevitch, M., Cassano, R., {et~al.} 2019, The Astrophysical Journal, 880, 70

\bibitem[{Giacintucci {et~al.}(2013)Giacintucci, Markevitch, Venturi, Clarke, Cassano, \& Mazzotta}]{Giacintucci_2013}
Giacintucci, S., Markevitch, M., Venturi, T., {et~al.} 2013, The Astrophysical Journal, 781, 9, \dodoi{10.1088/0004-637x/781/1/9}

\bibitem[{{Giacintucci} {et~al.}(2024){Giacintucci}, {Venturi}, {Markevitch}, {Brunetti}, {Clarke}, \& {Kale}}]{Giacintucci2024}
{Giacintucci}, S., {Venturi}, T., {Markevitch}, M., {et~al.} 2024, \apj, 961, 133, \dodoi{10.3847/1538-4357/ad12bc}

\bibitem[{Gitti {et~al.}(2004)Gitti, Brunetti, Feretti, \& Setti}]{gitti2004particle}
Gitti, M., Brunetti, G., Feretti, L., \& Setti, G. 2004, Astronomy \& Astrophysics, 417, 1

\bibitem[{{Gupta} {et~al.}(2017){Gupta}, {Ajithkumar}, {Kale}, {Nayak}, {Sabhapathy}, {Sureshkumar}, {Swami}, {Chengalur}, {Ghosh}, {Ishwara-Chandra}, {Joshi}, {Kanekar}, {Lal}, \& {Roy}}]{Gupta2017}
{Gupta}, Y., {Ajithkumar}, B., {Kale}, H.~S., {et~al.} 2017, Current Science, 113, 707, \dodoi{10.18520/cs/v113/i04/707-714}

\bibitem[{Hlavacek-Larrondo {et~al.}(2020)Hlavacek-Larrondo, Rhea, Webb, McDonald, Muzzin, Wilson, Finner, Valin, Bonaventura, Cooper, Fabian, Gendron-Marsolais, Jee, Lidman, Mezcua, Noble, Russell, Surace, Trudeau, \& Yee}]{Hlavacek_Larrondo_2020}
Hlavacek-Larrondo, J., Rhea, C.~L., Webb, T., {et~al.} 2020, The Astrophysical Journal, 898, L50, \dodoi{10.3847/2041-8213/ab9ca5}

\bibitem[{Hurley-Walker {et~al.}(2017)Hurley-Walker, Callingham, Hancock, Franzen, Hindson, Kapi{\'n}ska, Morgan, Offringa, Wayth, Wu, {et~al.}}]{hurley2017galactic}
Hurley-Walker, N., Callingham, J.~R., Hancock, P.~J., {et~al.} 2017, Monthly Notices of the Royal Astronomical Society, 464, 1146

\bibitem[{{Intema} {et~al.}(2017){Intema}, {Jagannathan}, {Mooley}, \& {Frail}}]{Intema2017}
{Intema}, H.~T., {Jagannathan}, P., {Mooley}, K.~P., \& {Frail}, D.~A. 2017, \aap, 598, A78, \dodoi{10.1051/0004-6361/201628536}

\bibitem[{{Jannuzi} \& {Dey}(1999)}]{Jannuzi1999}
{Jannuzi}, B.~T., \& {Dey}, A. 1999, in Astronomical Society of the Pacific Conference Series, Vol. 191, Photometric Redshifts and the Detection of High Redshift Galaxies, ed. R.~{Weymann}, L.~{Storrie-Lombardi}, M.~{Sawicki}, \& R.~{Brunner}, 111

\bibitem[{{Jonas} \& {MeerKAT Team}(2016)}]{Jonas2016}
{Jonas}, J., \& {MeerKAT Team}. 2016, in MeerKAT Science: On the Pathway to the SKA, 1, \dodoi{10.22323/1.277.0001}

\bibitem[{Keshet {et~al.}(2010)Keshet, Markevitch, Birnboim, \& Loeb}]{Keshet2010}
Keshet, U., Markevitch, M., Birnboim, Y., \& Loeb, A. 2010, Astrophys. J. Lett., 719, L74, \dodoi{10.1088/2041-8205/719/1/L74}

\bibitem[{Knowles {et~al.}(2019)Knowles, Baker, Bond, Gallardo, Gupta, Hilton, Hughes, Intema, L{\'{o} }pez-Caraballo, Moodley, Schmitt, Sievers, Sif{\'{o}}n, \& Wollack}]{Knowles_2019}
Knowles, K., Baker, A.~J., Bond, J.~R., {et~al.} 2019, Monthly Notices of the Royal Astronomical Society, 486, 1332, \dodoi{10.1093/mnras/stz823}

\bibitem[{Knowles {et~al.}(2022)Knowles, Cotton, Rudnick, Camilo, Goedhart, Deane, Ramatsoku, Bietenholz, Br{\"u}ggen, Button, {et~al.}}]{knowles2022meerkat}
Knowles, K., Cotton, W., Rudnick, L., {et~al.} 2022, Astronomy \& Astrophysics, 657, A56

\bibitem[{{Kondapally} {et~al.}(2021){Kondapally}, {Best}, {Hardcastle}, {Nisbet}, {Bonato}, {Sabater}, {Duncan}, {McCheyne}, {Cochrane}, {Bowler}, {Williams}, {Shimwell}, {Tasse}, {Croston}, {Goyal}, {Jamrozy}, {Jarvis}, {Mahatma}, {R{\"o}ttgering}, {Smith}, {Wo{\l}owska}, {Bondi}, {Brienza}, {Brown}, {Br{\"u}ggen}, {Chambers}, {Garrett}, {G{\"u}rkan}, {Huber}, {Kunert-Bajraszewska}, {Magnier}, {Mingo}, {Mostert}, {Nikiel-Wroczy{\'n}ski}, {O'Sullivan}, {Paladino}, {Ploeckinger}, {Prandoni}, {Rosenthal}, {Schwarz}, {Shulevski}, {Wagenveld}, \& {Wang}}]{Kondapally2021}
{Kondapally}, R., {Best}, P.~N., {Hardcastle}, M.~J., {et~al.} 2021, \aap, 648, A3, \dodoi{10.1051/0004-6361/202038813}

\bibitem[{Lindner {et~al.}(2014)Lindner, Baker, Hughes, Battaglia, Gupta, Knowles, Marriage, Menanteau, Moodley, Reese, \& Srianand}]{Lindner_2014}
Lindner, R.~R., Baker, A.~J., Hughes, J.~P., {et~al.} 2014, The Astrophysical Journal, 786, 49, \dodoi{10.1088/0004-637x/786/1/49}

\bibitem[{{Lockman} {et~al.}(1986){Lockman}, {Jahoda}, \& {McCammon}}]{Lockman1986}
{Lockman}, F.~J., {Jahoda}, K., \& {McCammon}, D. 1986, \apj, 302, 432, \dodoi{10.1086/164002}

\bibitem[{{Lusetti} {et~al.}(2024){Lusetti}, {Bonafede}, {Lovisari}, {Gitti}, {Ettori}, {Cassano}, {Riseley}, {Govoni}, {Br{\"u}ggen}, {Bruno}, {van Weeren}, {Botteon}, {Hoang}, {Gastaldello}, {Ignesti}, {Rossetti}, \& {Shimwell}}]{Lusetti2024}
{Lusetti}, G., {Bonafede}, A., {Lovisari}, L., {et~al.} 2024, \aap, 683, A132, \dodoi{10.1051/0004-6361/202347635}

\bibitem[{{Mazzotta} \& {Giacintucci}(2008)}]{Mazzotta2008}
{Mazzotta}, P., \& {Giacintucci}, S. 2008, \apjl, 675, L9, \dodoi{10.1086/529433}

\bibitem[{Morabito {et~al.}(2022)Morabito, Jackson, Mooney, {et~al.}}]{morabito21}
Morabito, L., Jackson, N., Mooney, S., {et~al.} 2022, A\&A, 658, A1, \dodoi{10.1051/0004-6361/202140649}

\bibitem[{Offringa {et~al.}(2013)Offringa, {de Bruyn}, Zaroubi, {et~al.}}]{Offringa13}
Offringa, A.~R., {de Bruyn}, A.~G., Zaroubi, S., {et~al.} 2013, A\&A, 549, A11, \dodoi{10.1051/0004-6361/201220293}

\bibitem[{Offringa {et~al.}(2015)Offringa, Wayth, Hurley-Walker, {et~al.}}]{Offringa15}
Offringa, A.~R., Wayth, R.~B., Hurley-Walker, N., {et~al.} 2015, PASA, 32, e008, \dodoi{10.1017/pasa.2015.7}

\bibitem[{{Oliver} {et~al.}(2000){Oliver}, {Rowan-Robinson}, {Alexander}, {Almaini}, {Balcells}, {Baker}, {Barcons}, {Barden}, {Bellas-Velidis}, {Cabrera-Guerra}, {Carballo}, {Cesarsky}, {Ciliegi}, {Clements}, {Crockett}, {Danese}, {Dapergolas}, {Drolias}, {Eaton}, {Efstathiou}, {Egami}, {Elbaz}, {Fadda}, {Fox}, {Franceschini}, {Genzel}, {Goldschmidt}, {Graham}, {Gonzalez-Serrano}, {Gonzalez-Solares}, {Granato}, {Gruppioni}, {Herbstmeier}, {H{\'e}raudeau}, {Joshi}, {Kontizas}, {Kontizas}, {Kotilainen}, {Kunze}, {La Franca}, {Lari}, {Lawrence}, {Lemke}, {Linden-V{\o}rnle}, {Mann}, {M{\'a}rquez}, {Masegosa}, {Mattila}, {McMahon}, {Miley}, {Missoulis}, {Mobasher}, {Morel}, {N{\o}rgaard-Nielsen}, {Omont}, {Papadopoulos}, {Perez-Fournon}, {Puget}, {Rigopoulou}, {Rocca-Volmerange}, {Serjeant}, {Silva}, {Sumner}, {Surace}, {Vaisanen}, {van der Werf}, {Verma}, {Vigroux}, {Villar-Martin}, \& {Willott}}]{Oliver2000}
{Oliver}, S., {Rowan-Robinson}, M., {Alexander}, D.~M., {et~al.} 2000, \mnras, 316, 749, \dodoi{10.1046/j.1365-8711.2000.03550.x}

\bibitem[{{Osinga} {et~al.}(2021){Osinga}, {van Weeren}, {Boxelaar}, {Brunetti}, {Botteon}, {Br{\"u}ggen}, {Shimwell}, {Bonafede}, {Best}, {Bonato}, {Cassano}, {Gastaldello}, {di Gennaro}, {Hardcastle}, {Mandal}, {Rossetti}, {R{\"o}ttgering}, {Sabater}, \& {Tasse}}]{Osinga_2021}
{Osinga}, E., {van Weeren}, R.~J., {Boxelaar}, J.~M., {et~al.} 2021, \aap, 648, A11, \dodoi{10.1051/0004-6361/202039076}

\bibitem[{{Perley} {et~al.}(2011){Perley}, {Chandler}, {Butler}, \& {Wrobel}}]{Perley2011}
{Perley}, R.~A., {Chandler}, C.~J., {Butler}, B.~J., \& {Wrobel}, J.~M. 2011, \apjl, 739, L1, \dodoi{10.1088/2041-8205/739/1/L1}

\bibitem[{{Pfrommer} \& {En{\ss}lin}(2003)}]{Pfrommer2003}
{Pfrommer}, C., \& {En{\ss}lin}, T.~A. 2003, \aap, 407, L73, \dodoi{10.1051/0004-6361:20031088}

\bibitem[{{Pfrommer} \& {En{\ss}lin}(2004{\natexlab{a}})}]{Pfrommer2004}
---. 2004{\natexlab{a}}, \aap, 413, 17, \dodoi{10.1051/0004-6361:20031464}

\bibitem[{{Pfrommer} \& {En{\ss}lin}(2004{\natexlab{b}})}]{Pfrommer2004_MEC}
---. 2004{\natexlab{b}}, \mnras, 352, 76, \dodoi{10.1111/j.1365-2966.2004.07900.x}

\bibitem[{{Pfrommer} {et~al.}(2008){Pfrommer}, {En{\ss}lin}, \& {Springel}}]{Pfrommer2008}
{Pfrommer}, C., {En{\ss}lin}, T.~A., \& {Springel}, V. 2008, \mnras, 385, 1211, \dodoi{10.1111/j.1365-2966.2008.12956.x}

\bibitem[{{Pfrommer} {et~al.}(2017){Pfrommer}, {Pakmor}, {Simpson}, \& {Springel}}]{Pfrommer2017}
{Pfrommer}, C., {Pakmor}, R., {Simpson}, C.~M., \& {Springel}, V. 2017, \apjl, 847, L13, \dodoi{10.3847/2041-8213/aa8bb1}

\bibitem[{{Prasow-{\'E}mond} {et~al.}(2020){Prasow-{\'E}mond}, {Hlavacek-Larrondo}, {Rhea}, {Latulippe}, {Gendron-Marsolais}, {Richard-Laferri{\`e}re}, {Sanders}, {Edge}, {Allen}, {Mantz}, \& {von der Linden}}]{PE2020}
{Prasow-{\'E}mond}, M., {Hlavacek-Larrondo}, J., {Rhea}, C.~L., {et~al.} 2020, \aj, 160, 103, \dodoi{10.3847/1538-3881/ab9ff3}

\bibitem[{{Quataert} \& {Hopkins}(2025)}]{Quataert2025}
{Quataert}, E., \& {Hopkins}, P.~F. 2025, arXiv e-prints, arXiv:2502.01753, \dodoi{10.48550/arXiv.2502.01753}

\bibitem[{Raja {et~al.}(2020)Raja, Rahaman, Datta, Burns, Alden, Intema, van Weeren, Hallman, Rapetti, \& Paul}]{Raja_2020}
Raja, R., Rahaman, M., Datta, A., {et~al.} 2020, The Astrophysical Journal, 889, 128, \dodoi{10.3847/1538-4357/ab620d}

\bibitem[{{Richard-Laferri{\`e}re} {et~al.}(2020){Richard-Laferri{\`e}re}, {Hlavacek-Larrondo}, {Nemmen}, {Rhea}, {Taylor}, {Prasow-{\'E}mond}, {Gendron-Marsolais}, {Latulippe}, {Edge}, {Fabian}, {Sanders}, {Hogan}, \& {Demontigny}}]{RL2020}
{Richard-Laferri{\`e}re}, A., {Hlavacek-Larrondo}, J., {Nemmen}, R.~S., {et~al.} 2020, \mnras, 499, 2934, \dodoi{10.1093/mnras/staa2877}

\bibitem[{{Ruszkowski} \& {Pfrommer}(2023)}]{Ruszkowski_2023}
{Ruszkowski}, M., \& {Pfrommer}, C. 2023, \aapr, 31, 4, \dodoi{10.1007/s00159-023-00149-2}

\bibitem[{{Rybicki} \& {Lightman}(1979)}]{Rybicki1979}
{Rybicki}, G.~B., \& {Lightman}, A.~P. 1979, {Radiative processes in astrophysics}

\bibitem[{{Sabater} {et~al.}(2021){Sabater}, {Best}, {Tasse}, {Hardcastle}, {Shimwell}, {Nisbet}, {Jelic}, {Callingham}, {R{\"o}ttgering}, {Bonato}, {Bondi}, {Ciardi}, {Cochrane}, {Jarvis}, {Kondapally}, {Koopmans}, {O'Sullivan}, {Prandoni}, {Schwarz}, {Smith}, {Wang}, {Williams}, \& {Zaroubi}}]{Sabater2021}
{Sabater}, J., {Best}, P.~N., {Tasse}, C., {et~al.} 2021, \aap, 648, A2, \dodoi{10.1051/0004-6361/202038828}

\bibitem[{{Savini} {et~al.}(2018){Savini}, {Bonafede}, {Br{\"u}ggen}, {van Weeren}, {Brunetti}, {Intema}, {Botteon}, {Shimwell}, {Wilber}, {Rafferty}, {Giacintucci}, {Cassano}, {Cuciti}, {de Gasperin}, {R{\"o}ttgering}, {Hoeft}, \& {White}}]{Savini2018}
{Savini}, F., {Bonafede}, A., {Br{\"u}ggen}, M., {et~al.} 2018, \mnras, 478, 2234, \dodoi{10.1093/mnras/sty1125}

\bibitem[{{Savini} {et~al.}(2019){Savini}, {Bonafede}, {Br{\"u}ggen}, {Rafferty}, {Shimwell}, {Botteon}, {Brunetti}, {Intema}, {Wilber}, {Cassano}, {Vazza}, {van Weeren}, {Cuciti}, {De Gasperin}, {R{\"o}ttgering}, {Sommer}, {B{\^\i}rzan}, \& {Drabent}}]{Savini2019}
---. 2019, \aap, 622, A24, \dodoi{10.1051/0004-6361/201833882}

\bibitem[{Shimwell {et~al.}(2017)Shimwell, R{\"o}ttgering, Best, Williams, Dijkema, de~Gasperin, Hardcastle, Heald, Hoang, Horneffer, {et~al.}}]{shimwell2017lofar}
Shimwell, T., R{\"o}ttgering, H., Best, P.~N., {et~al.} 2017, Astronomy \& Astrophysics, 598, A104

\bibitem[{{Shimwell} {et~al.}(2019){Shimwell}, {Tasse}, {Hardcastle}, {Mechev}, {Williams}, {Best}, {R{\"o}ttgering}, {Callingham}, {Dijkema}, {de Gasperin}, {Hoang}, {Hugo}, {Mirmont}, {Oonk}, {Prandoni}, {Rafferty}, {Sabater}, {Smirnov}, {van Weeren}, {White}, {Atemkeng}, {Bester}, {Bonnassieux}, {Br{\"u}ggen}, {Brunetti}, {Chy{\.z}y}, {Cochrane}, {Conway}, {Croston}, {Danezi}, {Duncan}, {Haverkorn}, {Heald}, {Iacobelli}, {Intema}, {Jackson}, {Jamrozy}, {Jarvis}, {Lakhoo}, {Mevius}, {Miley}, {Morabito}, {Morganti}, {Nisbet}, {Orr{\'u}}, {Perkins}, {Pizzo}, {Schrijvers}, {Smith}, {Vermeulen}, {Wise}, {Alegre}, {Bacon}, {van Bemmel}, {Beswick}, {Bonafede}, {Botteon}, {Bourke}, {Brienza}, {Calistro Rivera}, {Cassano}, {Clarke}, {Conselice}, {Dettmar}, {Drabent}, {Dumba}, {Emig}, {En{\ss}lin}, {Ferrari}, {Garrett}, {G{\'e}nova-Santos}, {Goyal}, {G{\"u}rkan}, {Hale}, {Harwood}, {Heesen}, {Hoeft}, {Horellou}, {Jackson}, {Kokotanekov}, {Kondapally}, {Kunert-Bajraszewska}, {Mahatma}, {Mahony}, {Mandal}, {McKean},
  {Merloni}, {Mingo}, {Miskolczi}, {Mooney}, {Nikiel-Wroczy{\'n}ski}, {O'Sullivan}, {Quinn}, {Reich}, {Roskowi{\'n}ski}, {Rowlinson}, {Savini}, {Saxena}, {Schwarz}, {Shulevski}, {Sridhar}, {Stacey}, {Urquhart}, {van der Wiel}, {Varenius}, {Webster}, \& {Wilber}}]{Shimwell_2019}
{Shimwell}, T.~W., {Tasse}, C., {Hardcastle}, M.~J., {et~al.} 2019, \aap, 622, A1, \dodoi{10.1051/0004-6361/201833559}

\bibitem[{{Shimwell} {et~al.}(2022){Shimwell}, {Hardcastle}, {Tasse}, {Best}, {R{\"o}ttgering}, {Williams}, {Botteon}, {Drabent}, {Mechev}, {Shulevski}, {van Weeren}, {Bester}, {Br{\"u}ggen}, {Brunetti}, {Callingham}, {Chy{\.z}y}, {Conway}, {Dijkema}, {Duncan}, {de Gasperin}, {Hale}, {Haverkorn}, {Hugo}, {Jackson}, {Mevius}, {Miley}, {Morabito}, {Morganti}, {Offringa}, {Oonk}, {Rafferty}, {Sabater}, {Smith}, {Schwarz}, {Smirnov}, {O'Sullivan}, {Vedantham}, {White}, {Albert}, {Alegre}, {Asabere}, {Bacon}, {Bonafede}, {Bonnassieux}, {Brienza}, {Bilicki}, {Bonato}, {Calistro Rivera}, {Cassano}, {Cochrane}, {Croston}, {Cuciti}, {Dallacasa}, {Danezi}, {Dettmar}, {Di Gennaro}, {Edler}, {En{\ss}lin}, {Emig}, {Franzen}, {Garc{\'\i}a-Vergara}, {Grange}, {G{\"u}rkan}, {Hajduk}, {Heald}, {Heesen}, {Hoang}, {Hoeft}, {Horellou}, {Iacobelli}, {Jamrozy}, {Jeli{\'c}}, {Kondapally}, {Kukreti}, {Kunert-Bajraszewska}, {Magliocchetti}, {Mahatma}, {Ma{\l}ek}, {Mandal}, {Massaro}, {Meyer-Zhao}, {Mingo}, {Mostert}, {Nair},
  {Nakoneczny}, {Nikiel-Wroczy{\'n}ski}, {Orr{\'u}}, {Pajdosz-{\'S}mierciak}, {Pasini}, {Prandoni}, {van Piggelen}, {Rajpurohit}, {Retana-Montenegro}, {Riseley}, {Rowlinson}, {Saxena}, {Schrijvers}, {Sweijen}, {Siewert}, {Timmerman}, {Vaccari}, {Vink}, {West}, {Wo{\l}owska}, {Zhang}, \& {Zheng}}]{shimwell2022}
{Shimwell}, T.~W., {Hardcastle}, M.~J., {Tasse}, C., {et~al.} 2022, \aap, 659, A1, \dodoi{10.1051/0004-6361/202142484}

\bibitem[{{Sikhosana} {et~al.}(2024){Sikhosana}, {Hilton}, {Bernardi}, {Kesebonye}, {Klutse}, {Knowles}, {Moodley}, {Mroczkowski}, {Partridge}, {Sif{\'o}n}, {Vargas}, \& {Wollack}}]{Sikhosona2024}
{Sikhosana}, S.~P., {Hilton}, M., {Bernardi}, G., {et~al.} 2024, arXiv e-prints, arXiv:2404.03944, \dodoi{10.48550/arXiv.2404.03944}

\bibitem[{{Tasse} {et~al.}(2021){Tasse}, {Shimwell}, {Hardcastle}, {O'Sullivan}, {van Weeren}, {Best}, {Bester}, {Hugo}, {Smirnov}, {Sabater}, {Calistro-Rivera}, {de Gasperin}, {Morabito}, {R{\"o}ttgering}, {Williams}, {Bonato}, {Bondi}, {Botteon}, {Br{\"u}ggen}, {Brunetti}, {Chy{\.z}y}, {Garrett}, {G{\"u}rkan}, {Jarvis}, {Kondapally}, {Mandal}, {Prandoni}, {Repetti}, {Retana-Montenegro}, {Schwarz}, {Shulevski}, \& {Wiaux}}]{Tasse2021}
{Tasse}, C., {Shimwell}, T., {Hardcastle}, M.~J., {et~al.} 2021, \aap, 648, A1, \dodoi{10.1051/0004-6361/202038804}

\bibitem[{{Tevlin} {et~al.}(2024){Tevlin}, {Berlok}, {Pfrommer}, {Talbot}, {Whittingham}, {Puchwein}, {Pakmor}, {Weinberger}, \& {Springel}}]{Tevlin_2024}
{Tevlin}, L., {Berlok}, T., {Pfrommer}, C., {et~al.} 2024, arXiv e-prints, arXiv:2411.00103, \dodoi{10.48550/arXiv.2411.00103}

\bibitem[{Timmerman {et~al.}(2021)Timmerman, van Weeren, McDonald, Ignesti, McNamara, Hlavacek-Larrondo, \& Röttgering}]{Timmerman_2021}
Timmerman, R., van Weeren, R.~J., McDonald, M., {et~al.} 2021, Astronomy \& Astrophysics, 646, A38, \dodoi{10.1051/0004-6361/202039075}

\bibitem[{{Tolman}(1934)}]{Tolman1934}
{Tolman}, R.~C. 1934, {Relativity, Thermodynamics, and Cosmology}

\bibitem[{Trudeau {et~al.}(2019)Trudeau, Webb, Hlavacek-Larrondo, Noble, Gendron-Marsolais, Lidman, Mezcua, Muzzin, Wilson, \& Yee}]{trudeau_multiwavelength_2019}
Trudeau, A., Webb, T., Hlavacek-Larrondo, J., {et~al.} 2019, Monthly Notices of the Royal Astronomical Society, 487, 1210, \dodoi{10.1093/mnras/stz1364}

\bibitem[{{van Haarlem} {et~al.}(2013){van Haarlem}, {Wise}, {Gunst}, {Heald}, {McKean}, {Hessels}, {de Bruyn}, {Nijboer}, {Swinbank}, {Fallows}, {Brentjens}, {Nelles}, {Beck}, {Falcke}, {Fender}, {H{\"o}randel}, {Koopmans}, {Mann}, {Miley}, {R{\"o}ttgering}, {Stappers}, {Wijers}, {Zaroubi}, {van den Akker}, {Alexov}, {Anderson}, {Anderson}, {van Ardenne}, {Arts}, {Asgekar}, {Avruch}, {Batejat}, {B{\"a}hren}, {Bell}, {Bell}, {van Bemmel}, {Bennema}, {Bentum}, {Bernardi}, {Best}, {B{\^\i}rzan}, {Bonafede}, {Boonstra}, {Braun}, {Bregman}, {Breitling}, {van de Brink}, {Broderick}, {Broekema}, {Brouw}, {Br{\"u}ggen}, {Butcher}, {van Cappellen}, {Ciardi}, {Coenen}, {Conway}, {Coolen}, {Corstanje}, {Damstra}, {Davies}, {Deller}, {Dettmar}, {van Diepen}, {Dijkstra}, {Donker}, {Doorduin}, {Dromer}, {Drost}, {van Duin}, {Eisl{\"o}ffel}, {van Enst}, {Ferrari}, {Frieswijk}, {Gankema}, {Garrett}, {de Gasperin}, {Gerbers}, {de Geus}, {Grie{\ss}meier}, {Grit}, {Gruppen}, {Hamaker}, {Hassall}, {Hoeft}, {Holties},
  {Horneffer}, {van der Horst}, {van Houwelingen}, {Huijgen}, {Iacobelli}, {Intema}, {Jackson}, {Jelic}, {de Jong}, {Juette}, {Kant}, {Karastergiou}, {Koers}, {Kollen}, {Kondratiev}, {Kooistra}, {Koopman}, {Koster}, {Kuniyoshi}, {Kramer}, {Kuper}, {Lambropoulos}, {Law}, {van Leeuwen}, {Lemaitre}, {Loose}, {Maat}, {Macario}, {Markoff}, {Masters}, {McFadden}, {McKay-Bukowski}, {Meijering}, {Meulman}, {Mevius}, {Middelberg}, {Millenaar}, {Miller-Jones}, {Mohan}, {Mol}, {Morawietz}, {Morganti}, {Mulcahy}, {Mulder}, {Munk}, {Nieuwenhuis}, {van Nieuwpoort}, {Noordam}, {Norden}, {Noutsos}, {Offringa}, {Olofsson}, {Omar}, {Orr{\'u}}, {Overeem}, {Paas}, {Pandey-Pommier}, {Pandey}, {Pizzo}, {Polatidis}, {Rafferty}, {Rawlings}, {Reich}, {de Reijer}, {Reitsma}, {Renting}, {Riemers}, {Rol}, {Romein}, {Roosjen}, {Ruiter}, {Scaife}, {van der Schaaf}, {Scheers}, {Schellart}, {Schoenmakers}, {Schoonderbeek}, {Serylak}, {Shulevski}, {Sluman}, {Smirnov}, {Sobey}, {Spreeuw}, {Steinmetz}, {Sterks}, {Stiepel}, {Stuurwold},
  {Tagger}, {Tang}, {Tasse}, {Thomas}, {Thoudam}, {Toribio}, {van der Tol}, {Usov}, {van Veelen}, {van der Veen}, {ter Veen}, {Verbiest}, {Vermeulen}, {Vermaas}, {Vocks}, {Vogt}, {de Vos}, {van der Wal}, {van Weeren}, {Weggemans}, {Weltevrede}, {White}, {Wijnholds}, {Wilhelmsson}, {Wucknitz}, {Yatawatta}, {Zarka}, {Zensus}, \& {van Zwieten}}]{vanHaarlem2013}
{van Haarlem}, M.~P., {Wise}, M.~W., {Gunst}, A.~W., {et~al.} 2013, \aap, 556, A2, \dodoi{10.1051/0004-6361/201220873}

\bibitem[{{van Weeren} {et~al.}(2019){van Weeren}, {de Gasperin}, {Akamatsu}, {Br{\"u}ggen}, {Feretti}, {Kang}, {Stroe}, \& {Zandanel}}]{van_Weeren_2019}
{van Weeren}, R.~J., {de Gasperin}, F., {Akamatsu}, H., {et~al.} 2019, \ssr, 215, 16, \dodoi{10.1007/s11214-019-0584-z}

\bibitem[{van Weeren {et~al.}(2014{\natexlab{a}})van Weeren, Intema, Lal, Bonafede, Jones, Forman, Röttgering, Brüggen, Stroe, Hoeft, Nuza, \& de~Gasperin}]{van_Weeren_2014}
van Weeren, R.~J., Intema, H.~T., Lal, D.~V., {et~al.} 2014{\natexlab{a}}, The Astrophysical Journal, 781, L32, \dodoi{10.1088/2041-8205/781/2/l32}

\bibitem[{van Weeren {et~al.}(2014{\natexlab{b}})van Weeren, Intema, Lal, Andrade-Santos, Brüggen, de~Gasperin, Forman, Hoeft, Jones, Nuza, Röttgering, \& Stroe}]{van_Weeren_2014_phoenix}
---. 2014{\natexlab{b}}, The Astrophysical Journal, 786, L17, \dodoi{10.1088/2041-8205/786/2/l17}

\bibitem[{{van Weeren} {et~al.}(2016){van Weeren}, {Williams}, {Hardcastle}, {Shimwell}, {Rafferty}, {Sabater}, {Heald}, {Sridhar}, {Dijkema}, {Brunetti}, {Br{\"u}ggen}, {Andrade-Santos}, {Ogrean}, {R{\"o}ttgering}, {Dawson}, {Forman}, {de Gasperin}, {Jones}, {Miley}, {Rudnick}, {Sarazin}, {Bonafede}, {Best}, {B{\^\i}rzan}, {Cassano}, {Chy{\.z}y}, {Croston}, {Ensslin}, {Ferrari}, {Hoeft}, {Horellou}, {Jarvis}, {Kraft}, {Mevius}, {Intema}, {Murray}, {Orr{\'u}}, {Pizzo}, {Simionescu}, {Stroe}, {van der Tol}, \& {White}}]{vanWeeren2016}
{van Weeren}, R.~J., {Williams}, W.~L., {Hardcastle}, M.~J., {et~al.} 2016, \apjs, 223, 2, \dodoi{10.3847/0067-0049/223/1/2}

\bibitem[{{van Weeren} {et~al.}(2024){van Weeren}, {Timmerman}, {Vaidya}, {Gendron-Marsolais}, {Botteon}, {Roberts}, {Hlavacek-Larrondo}, {Bonafede}, {Br{\"u}ggen}, {Brunetti}, {Cassano}, {Cuciti}, {Edge}, {Gastaldello}, {Groeneveld}, \& {Shimwell}}]{vanweeren2024}
{van Weeren}, R.~J., {Timmerman}, R., {Vaidya}, V., {et~al.} 2024, arXiv e-prints, arXiv:2410.02863, \dodoi{10.48550/arXiv.2410.02863}

\bibitem[{Wayth {et~al.}(2015)Wayth, Lenc, Bell, Callingham, Dwarakanath, Franzen, Gaensler, Hancock, Hindson, Hurley-Walker, {et~al.}}]{wayth2015gleam}
Wayth, R., Lenc, E., Bell, M., {et~al.} 2015, Publications of the Astronomical Society of Australia, 32, e025

\bibitem[{Webb {et~al.}(2015{\natexlab{a}})Webb, Noble, DeGroot, Wilson, Muzzin, Bonaventura, Cooper, Delahaye, Foltz, Lidman, Surace, Yee, Chapman, Dunne, Geach, Hayden, Hildebrandt, Huang, Pope, Smith, Perlmutter, \& Tudorica}]{webb_extreme_2015}
Webb, T., Noble, A., DeGroot, A., {et~al.} 2015{\natexlab{a}}, The Astrophysical Journal, 809, 173, \dodoi{10.1088/0004-637X/809/2/173}

\bibitem[{Webb {et~al.}(2015{\natexlab{b}})Webb, Muzzin, Noble, Bonaventura, Geach, Hezevah, Lidman, Wilson, Yee, Surace, \& Shupe}]{webb_star_2015}
Webb, T. M.~A., Muzzin, A., Noble, A., {et~al.} 2015{\natexlab{b}}, The Astrophysical Journal, 814, 96, \dodoi{10.1088/0004-637X/814/2/96}

\bibitem[{Webb {et~al.}(2017)Webb, Lowenthal, Yun, Noble, Muzzin, Wilson, Yee, Cybulski, Aretxaga, \& Hughes}]{webb_detection_2017}
Webb, T. M.~A., Lowenthal, J., Yun, M., {et~al.} 2017, The Astrophysical Journal, 844, L17, \dodoi{10.3847/2041-8213/aa7749}

\bibitem[{{Weinberger} {et~al.}(2017){Weinberger}, {Springel}, {Hernquist}, {Pillepich}, {Marinacci}, {Pakmor}, {Nelson}, {Genel}, {Vogelsberger}, {Naiman}, \& {Torrey}}]{Weinberger2017}
{Weinberger}, R., {Springel}, V., {Hernquist}, L., {et~al.} 2017, \mnras, 465, 3291, \dodoi{10.1093/mnras/stw2944}

\bibitem[{{Williams} {et~al.}(2016){Williams}, {van Weeren}, {R{\"o}ttgering}, {Best}, {Dijkema}, {de Gasperin}, {Hardcastle}, {Heald}, {Prandoni}, {Sabater}, {Shimwell}, {Tasse}, {van Bemmel}, {Br{\"u}ggen}, {Brunetti}, {Conway}, {En{\ss}lin}, {Engels}, {Falcke}, {Ferrari}, {Haverkorn}, {Jackson}, {Jarvis}, {Kapi{\'n}ska}, {Mahony}, {Miley}, {Morabito}, {Morganti}, {Orr{\'u}}, {Retana-Montenegro}, {Sridhar}, {Toribio}, {White}, {Wise}, \& {Zwart}}]{Williams2016}
{Williams}, W.~L., {van Weeren}, R.~J., {R{\"o}ttgering}, H.~J.~A., {et~al.} 2016, \mnras, 460, 2385, \dodoi{10.1093/mnras/stw1056}

\bibitem[{{Zhang} {et~al.}(2023){Zhang}, {Simionescu}, {Gastaldello}, {Eckert}, {Camillini}, {Natale}, {Rossetti}, {Brunetti}, {Akamatsu}, {Botteon}, {Cassano}, {Cuciti}, {Bruno}, {Shimwell}, {Jones}, {Kaastra}, {Ettori}, {Br{\"u}ggen}, {de Gasperin}, {Drabent}, {van Weeren}, \& {R{\"o}ttgering}}]{Zhang2023}
{Zhang}, X., {Simionescu}, A., {Gastaldello}, F., {et~al.} 2023, \aap, 672, A42, \dodoi{10.1051/0004-6361/202244761}

\end{thebibliography}
\bibliographystyle{aasjournal}

\end{document}